\documentclass[12pt]{spieman}  % 12pt font required by SPIE;
\usepackage{amsmath,amsfonts,amssymb}
\let\oldAA\AA
\renewcommand{\AA}{\text{\normalfont\oldAA}} % for angstrom units
\usepackage{graphicx}
\usepackage{placeins}
\usepackage{setspace}
\usepackage{tocloft}
\usepackage{array,tabularx}
\usepackage{array}
\usepackage{booktabs}
\usepackage{multirow}% http://ctan.org/pkg/multirow
\usepackage{hhline}% http://ctan.org/pkg/hhline
\newenvironment{conditions}
  {\par\vspace{\abovedisplayskip}\noindent\begin{tabular}{>{$}l<{$} @{${}-{}$} l}}
  {\end{tabular}\par\vspace{\belowdisplayskip}}
  
\title{Evaluation of Digital Micromirror Devices for use in space-based Multi-Object Spectrometer application}

\author[a,*]{Anton Travinsky}
\author[a]{Dmitry Vorobiev}
\author[a]{Zoran Ninkov}
\author[b]{Alan Raisanen}
\author[c]{Manuel A. Quijada}
\author[d]{Stephen A. Smee}
\author[c]{Jonathan A. Pellish}
\author[c]{Tim Schwartz}
\author[e]{Massimo Robberto}
\author[c]{Sara Heap}
\author[f]{Devin Conley}
\author[f]{Carlos Benavides}
\author[f]{Nicholas Garcia}
\author[f]{Zach Bredl}
\author[f]{Sebastian Yllanes}

\affil[a]{Rochester Institute of Technology, Chester F. Carlson Center for Imaging Science, 54 Lomb Memorial Drive, Rochester, NY 14623 USA}
\affil[b]{Rochester Institute of Technology, Department of Manufacturing and Mechanical Engineering Technology, 78 Lomb Memorial Drive, Rochester, NY 14623 USA}
\affil[c]{NASA Goddard Space Flight Center, 8800 Greenbelt Rd, Greenbelt, MD 20771 USA}
\affil[e]{Space Telescope Sciences Institute, 3700 San Martin Dr., Baltimore, MD 21218 USA}
\affil[d]{Department of Physics and Astronomy, Johns Hopkins University, 3701 San Martin Drive, Baltimore, MD 21218 USA}
\affil[f]{Department of Mechanical Engineering, Johns Hopkins University, 3400 North Charles Street, Baltimore, MD 21218 USA}

\cftpagenumbersoff{figure}
\cftpagenumbersoff{table} 
\begin{document} 
\maketitle

\begin{abstract}
The astronomical community continues to be interested in suitable programmable slit masks for use in  multi-object spectrometers (MOSs) on space missions. There have been ground-based MOS utilizing digital micromirror devices (DMDs) and they have proven to be highly accurate and reliable instruments. This paper summarizes the results of a continuing study to investigate the performance of DMDs under conditions associated with space deployment.  This includes the response of DMDs to radiation, to the vibration and mechanical shock loads associated with launch, and the operability of DMD under cryogenic temperatures. The optical contrast ratio and a study of the long-term reflectance of a bare device have also been investigated. The results of the radiation testing demonstrate that DMDs in orbit would experience negligible heavy-ion induced single event upset (SEU) rate burden; we predict SEU rate of 5.6 micromirrors per 24 hours. Vibration and mechanical shock testing was performed according to the NASA General Environmental Verification Standard (GEVS); no mirrors failed in the devices tested. The results of low temperature testing suggest that DMDs are not affected by the thermal load and operate smoothly at  temperatures at least as low as 78 K. The reflectivity of a bare DMD did not measurably change even after being exposed to ambient conditions over a period of 13 months. The measured contrast ratio (\emph{on state} vs \emph{off state} of the DMD micromirrors) was greater than 6000:1 when illuminated with an $f/4$ optical beam. Overall, DMDs are extremely robust and promise to provide a reliable alternative to micro shutter arrays (MSA) to be used in space as remotely programmable slit masks for MOS design.   
\end{abstract}

% Include a list of up to six keywords after the abstract

% Include email contact information for corresponding author
{\noindent \footnotesize\textbf{*}Corresponding author, E-mail: \linkable{anton.travinsky@mail.rit.edu} }

\begin{spacing}{1}
%%%%%%%%%%%%%%%%%%%%%%%%%%%%%%%%%%%%%%%%%%%%%%%%%
% Introduction
%%%%%%%%%%%%%%%%%%%%%%%%%%%%%%%%%%%%%%%%%%%%%%%%%
\section{Introduction}
\label{sect:intro} 
%---------------------------------------------------------------------------------------------------------------
% DMDs
%----------------------------------------------------------------------------------------------------------------
\subsection{Digital Micromirror Devices}
\label{subsec:Intro:DMDs}
\begin{figure}[htb]
\centering
\includegraphics[width=\textwidth]{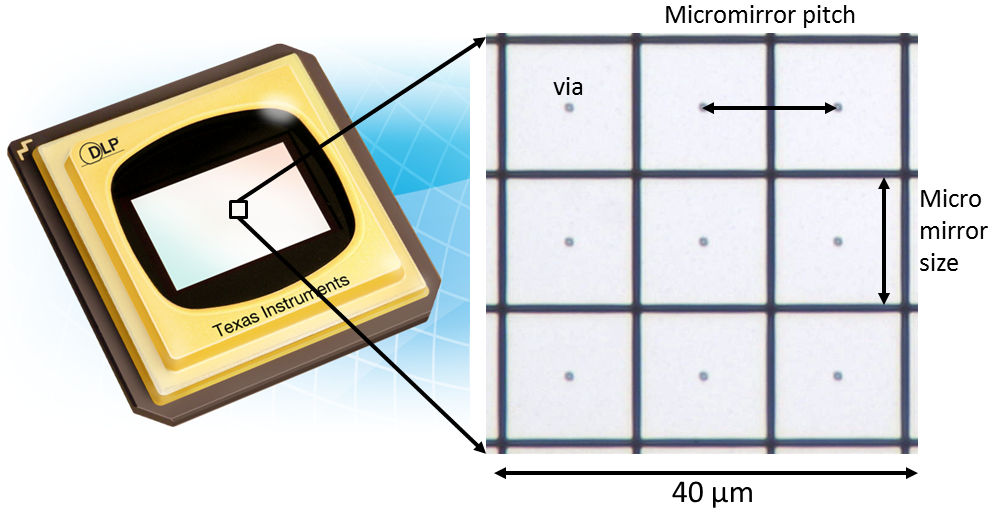} 
\setlength{\abovecaptionskip}{12pt}
\caption 
{ \label{fig:DMD}
TI\textsuperscript{\texttrademark}-packaged DMD in a hermetic case (\textit{left}, from \url{www.ti.com}) and microscopic image of a 3$\times$3 micromirror pixels (\textit{right}). Micromirror pitch is the distance between the centers of two neighboring mirrors, micromirror size is the linear dimension of one mirror, and a via is a small non-reflective part in the center on each micromirror, through which the mirrors are mounted onto the underlying hinge structure.} 
\end{figure}

\begin{figure}[htb]
\centering
\includegraphics[width=\textwidth]{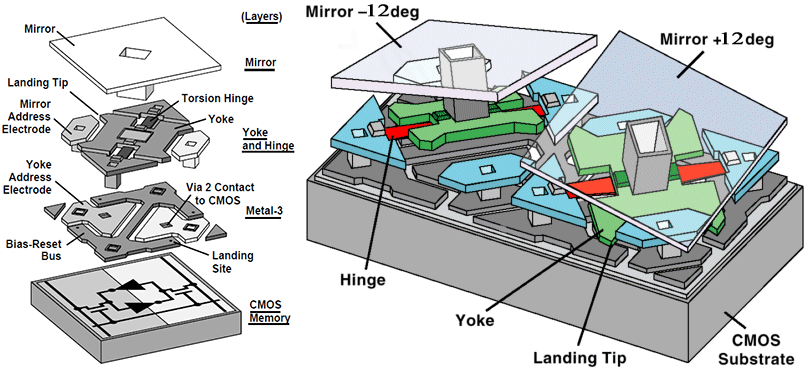} 
\setlength{\abovecaptionskip}{12pt}
\caption 
{ \label{fig:DMDschematics} Schematic of DMD pixels. \textit{(left hand side)} Exploded view of a single DMD pixel consisting of the micromirror itself, two electrode layers, and a complementary metal–oxide–semiconductor (CMOS) memory layer. \textit{(right hand side)} Two DMD pixels tilted into opposite operational states ($\pm$ 12$^{\circ}$).}
\end{figure}

A digital micromirror device (DMD) (Figure \ref{fig:DMD}), is a binary light modulator, which consists of a rectangular array of square micromirrors that can be tilted around their diagonals into either one of two stable positions (Figure \ref{fig:DMDschematics}). When the device is not powered the micromirrors lie flat, meaning the normal to the surface of each micromirror is aligned with the optical axis of the device. When the DMD is powered on, the micromirrors can be individually addressed and rotated into either one of their two fixed operating position. Those positions being the mirror tilted \(\pm\)12\(^{\circ}\) from the normal to the surface of the device, sometimes referred to as the \emph{on state} and the \emph{off state}. DMDs are the core of the Digital Light Processing (DLP\textsuperscript{\textregistered}) technology, patented by Texas Instruments Inc. (TI\textsuperscript{\texttrademark}).\\

The investing of several decades of research and development effort \cite{Hornbeck1993} into DMDs by TI\textsuperscript{\texttrademark} has resulted in lightweight, compact, and extremely reliable devices. In a DMD-based projection system image quality does not degrade over time\cite{Douglass2003}, there are practically no mechanical hinge failures or stuck mirrors\cite{Douglass2003,Zamkotsian2011_Successful_evaluation_for_space_applications_of_the_2048×1080_DMD,FourspringProtonRadTestingJournal,SontheimerHinge}.  Earlier reports have indicated that DMDs are tolerant to low temperatures\cite{Zamkotsian2011_Successful_evaluation_for_space_applications_of_the_2048×1080_DMD,FourspringPhD_Thesis}, and are not affected by vibration and mechanical shock\cite{Zamkotsian2011_Successful_evaluation_for_space_applications_of_the_2048×1080_DMD,MacKenty2003}. There is added testing required for space deployed instruments beyond that routinely done for terrestrial deployment. These included evaluation for the effects of radiation, thermal cycling, thermal shock, vibration and mechanical shock at launch, heat shielding, and operation in vacuum\cite{Shea2009}. Recently we performed and reported the results of proton radiation testing\cite{FourspringProtonRadTestingJournal}, thermal cycling calculations\cite{FourspringPhD_Thesis}, initial heavy ion radiation testing\cite{Travinsky_RadTestI_proceedings,Travinsky_Rad_test_Journal_I}, and vibration and mechanical shock testing\cite{Vorobiev_Vibration_Shock}. Previous results of proton radiation testing suggest that  proton fluxes such as those to be found in an orbit comparable to the second Lagrangian point (L2, interplanetary space), will not significantly affect the performance of DMDs under standard shielding\cite{FourspringProtonRadTestingJournal}.\\ 

\begin{figure}
\centering
\includegraphics[width=\textwidth]{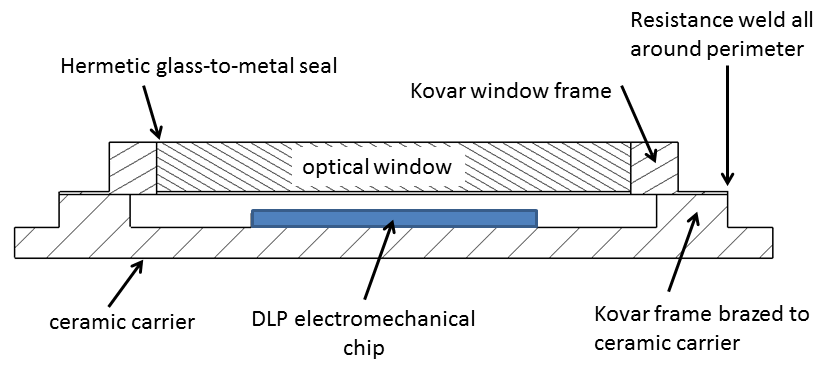} 
\setlength{\abovecaptionskip}{12pt}
\caption 
{ \label{fig:DMDpackage}
Cross section of a DMD package as provided by TI\textsuperscript{\texttrademark}\cite{Travinsky_Rad_test_Journal_I}. Not illustrated are two small getter pellets, which help maintain a low contamination environment inside the package.} 
\end{figure}

DMDs can differ in wavelength regime, number of micromirrors in the array, micromirror size, distance between the centers of two neighboring micromirrors (pitch) (Figure \ref{fig:DMD}), and micromirror tip angle. The gaps between the micromirrors are necessary to allow for rotation between the two operational stages. There is also a small non-reflective patch in the center of each micromirror - via. The via provides mechanical support for the micromirror and conducts electrical signals to the micromirror. Fill-factor signifies the ratio between the ``useful'' (reflective) area of a DMD and the complete area of the array, while the latter includes the gaps and the vias. Tests described in this paper were done with 0.7\textsuperscript{$\prime \prime$} 1024$\times$768 eXtended Graphics Array (XGA) DMDs. These devices have a micromirror tilt angle of 12$^{\circ}$, micromirror pitch of 13.68 \(\mu\)m, and fill-factor of 92\%\cite{DMDdatasheet}. The DMDs were controlled using the formatter board supplied by Digital Light Innovations (DLi) as a part of their DLP Discovery\textsuperscript{TM} D4100 kit\cite{DLi4100}. This model of DMDs is supplied with a hermetically sealed package, which consists of a ceramic carrier, kovar frame, and a borosilicate window (Corning 7056 glass, see Figure \ref{fig:DMDpackage}), and is filled with an inert gas. We evaluated and here report on the performance of these current-generation DMDs under the extreme conditions associated with space deployment.
\\
%------------------------------------------------------------------------
% MOS
%-------------------------------------------------------------------------
\subsection{Space-suitable DMD-based multi object spectrometers}
\label{subsec:Intro:MOS}
One particular type of scientific instrumentation, where DMDs can be extremely useful are multi object spectrometers (MOS)\cite{Robberto2009,Canonica2013}. Currently there are two predominant approaches to ground-based astronomical MOS, those being, using optical fibers (e.g. Hydra\cite{Barden1995}), or fabricated slit masks\cite{FourspringProtonRadTestingJournal}. Hydra is essentially a robot that moves around the focal plane of a telescope and plugs optical fibers into locations of interest. The fibers transfer the light to the spectrometer. The disadvantages of this technology, utilized, for instance, by the Subaru telescope\cite{Kimura2010}, are inherent complexity, limitation to a minimal slit width, and time of re-positioning of optical fibers (on the order of one hour). The second approach, slit masks, has been adopted by both Gemini telescopes in an instrument named “Gemini Multi Object Spectrograph” (GMOS\cite{Davies1997}). This technology requires capturing the image of the observation field and then machining a mask with rectangular holes (slits) in the locations on the focal plane corresponding to objects of interest. The purpose of this mask is to pass only the light from the objects of interest and nothing else. Masks must be produced weeks in advance with extreme precision. Both technologies, fibers and slit masks, are not suitable to use for space-based instruments.\\

%The possibility of using a solid state device that generates the needed randomly addressable re-programmable slits is very attractive to the astronomical community. Alternate mask generation approaches in MOS system relies on either bulky technology (e.g. fibers), non-real time approaches (e..g. laser manufactured slits in a plate) or lower yield transparent slit devices (e.g. microshutters). Some of these approaches are clearly not suitable for a space deployment.\\

\begin{figure}
\centering
\includegraphics[width=\textwidth]{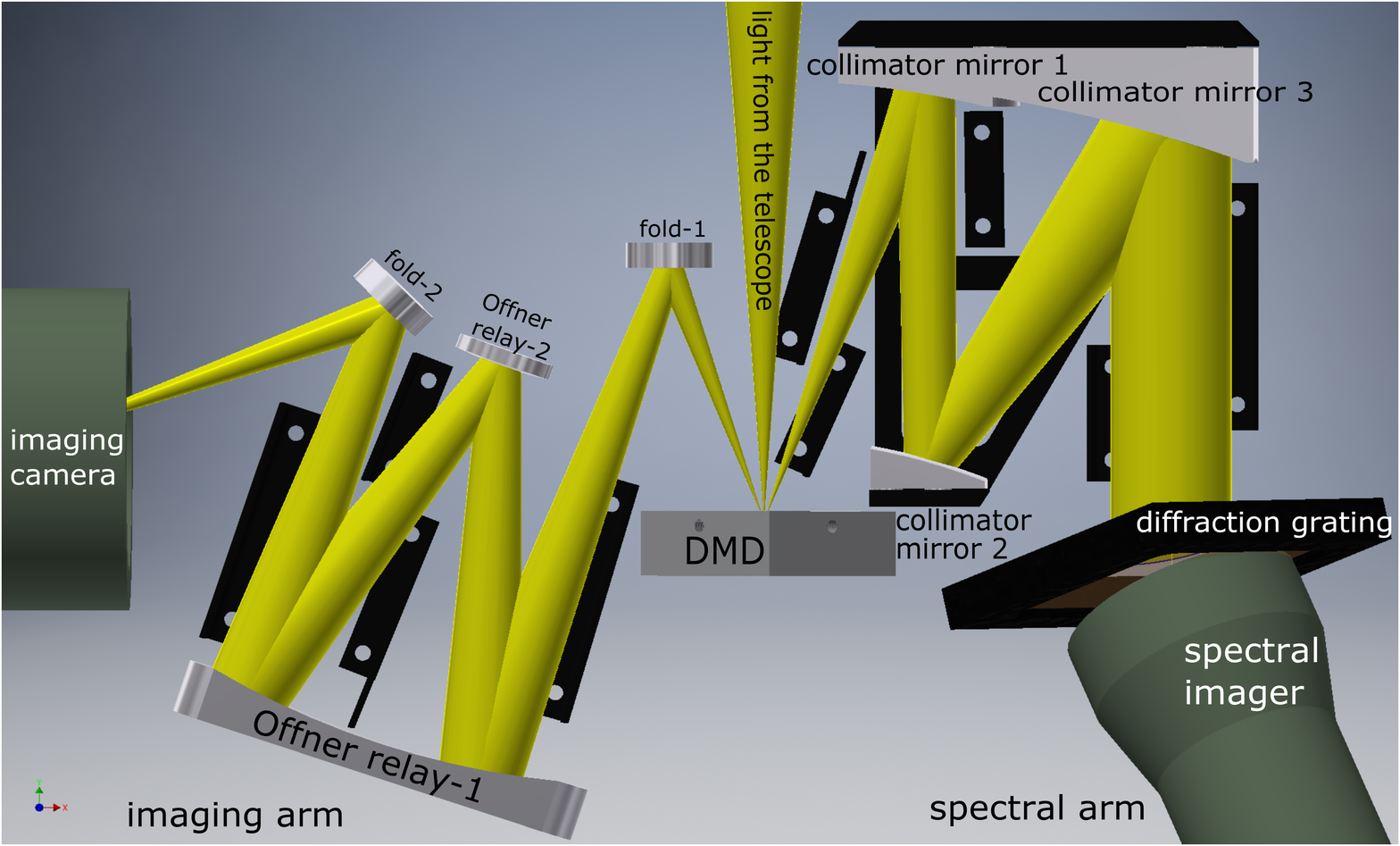} 
\setlength{\abovecaptionskip}{12pt}
\caption 
{ \label{fig:ritmos}
A typical layout of a DMD-based MOS such as found in RITMOS \cite{Kearney1998}. A DMD is located in the focal plane of a telescope. If all the micromirrors are turned towards the first folding mirror (marked as ``fold-1'') (the DMD micromirrors are in the \emph{off state}), the light from the telescope is reflected towards the imaging arm and the 2-mirror Offner imaging relay\cite{Offner1963} creates an image of the surface of the DMD on the imaging camera. If now the micromirrors corresponding to the objects of interest in the field of view are turned towards the spectral arm (into the \emph{on state}), the light from these locations in the field goes through the 3-mirror collimator, and a transmissive diffraction grating disperses it into spectra of these objects. The spectra are then acquired by the spectral imager.} 
\end{figure}

A typical layout of a DMD-based MOS such as found in RITMOS\cite{Kearney1998} includes two optical paths (Figure \ref{fig:ritmos}) - the imaging arm and the spectral arm. The DMD is placed at the focal plane of a telescope in such a way that the light from the telescope reaches the DMD at normal incidence. Initially by tilting the micromirrors into the \emph{off state} operating position, towards the imaging arm, one can obtain an image of the focal plane of the telescope on an imaging sensor. Then the micromirrors corresponding to objects of interest in the image can be tilted into the \emph{on state}, towards the spectral arm, where the light from each slit so formed by the micromirrors is dispersed onto a focal plane array. This way spectra of the multiple selected objects of interest can be integrated and recorded simultaneously. Also, the inherent ability to re-program the DMD mask pattern in just a few seconds \cite{Kearney1998,SAMOS} permits rapid changes to the targets being observed and to the planned observations if necessary.\\
\\

In the early stages of planning the James Webb Space Telescope (JWST) mission in the late 1990s, transmissive (micro-shutter arrays (MSAs))\cite{MoseleyMSA1999} and reflective (micromirror arrays (MMAs))\cite{MacKentyMOSusingMMA,MacKentyPrePhaseA} micro-electro-mechanical (MEMS) devices were investigated by the National Aeronautics and Space Administration (NASA) for possible use as programmable masks in space-based MOSs\cite{Robberto2009}. Although TI DMDs were already commercially available at the time, they were not considered because of the requirement for a single micromirror to match the GWST plate scale (100 $\times$ 100 $\mu$m), the required ability to operate at 30K, and the requirement for a defect and failure tendency towards the \emph{off state}. Instead, NASA supported the development of MMAs, which, like DMDs, are arrays of micromirrors that can be addressed individually and tilted into \emph{on state} or \emph{off state}, but the size of each micromirror in MMAs is about 100$\mu$m\cite{MacKentyPrePhaseA}. MMAs were developed both internally at NASA Goddard Space Flight Center (GSFC) \cite{MacKentyPrePhaseA} and externally at Sandia National Laboratories\cite{Garcia1}. Both MMAs and MSAs reached the Pre-Phase A study level\cite{MacKentyPrePhaseA,MoseleyMSA1999,SandiaNGSTpressRelease}, but the MSAs were picked over MMAs due to the latter's inability to achieve the desired contrast ratio at the time\cite{Robberto2009}.

Since the 1990's the DMD technology has advanced significantly and today’s DMDs feature a larger micromirror tilt angle, a smaller via, an improved micromirror surface coating, and more pixels. Studies using digital micromirror devices for MOS over the last two decades have been very successful\cite{Zamkotsian_Batman_Journal,Zamkotsian2011_Successful_evaluation_for_space_applications_of_the_2048×1080_DMD,Zamkotsian_Batman_Flies_proceedings,Zamkotsian2015_DMD-based_programmable_wide_field_spectrograph_for_Earth_Observation,Zamkotsian2014_Micromirror_arrays_for_spectroscopy_in_space,Kearney1998}. DMD-based MOS for use on ground telescopes have already been built and tested: Rochester Institute of Technology Multi-Object Spectrometer (RITMOS)\cite{Kearney1998,Meyer2004}, InfraRed Multi-Object Spectrometer (IRMOS)\cite{MacKenty2003,Winsor2000,Connelly2003}, and the most recent - BATMAN\cite{Zamkotsian_Batman_Flies_proceedings,Zamkotsian_Batman_Journal}. All of these studies and previous work suggests that DMDs represent a very promising technology for astronomy and remote sensing\cite{Uzkent2013,Uzkent2016} applications. This paper reports on a variety of testing designed to specifically evaluate how feasible it is to utilize commercial DMDs in space applications.
%%%%%%%%%%%%%%%%%%%%%%%%%%%%%%%%%%%%%%%%%%%%%%%%%
% radiation testing
%%%%%%%%%%%%%%%%%%%%%%%%%%%%%%%%%%%%%%%%%%%%%%%%%
\section{Radiation testing}
%-----------------------------------------------------------------------------
% Background Radiation 
%-----------------------------------------------------------------------------
\subsection{Background}
\label{subsec:RadtestBackground}
The operation of an electronic device on Earth differ from that in space in many ways, one of which is the response of the device to higher radiation level. We reported on the results of proton radiation testing\cite{FourspringProtonRadTestingJournal} in 2013 and heavy ion radiation experiment\cite{Travinsky_RadTestI_proceedings} in 2015. The outcome of the proton radiation testing\cite{FourspringProtonRadTestingJournal} suggests that the proton fluxes in an orbit comparable to the second Lagrange point (L2) in interplanetary space will not significantly affect the performance of DMD. Our previous results from heavy-ion radiation test\cite{Travinsky_Rad_test_Journal_I} support this conclusion. Based on these results we also predicted the heavy-ion induced in-orbit micromirror upset rate to be 10 micromirrors in 24 hours on orbit (less than 1 micromirror every two hours) for a 1 megapixel (MP) device\cite{Travinsky_Rad_test_Journal_I}.  This is negligible because in typical operation the mask pattern would be reset periodically over times less than one hour. Howevever these results also had wide 95\% confidence bounds which called for some healthy skepticism about the accuracy of the prediction. For this reason additional radiation testing experiments were undertaken that concentrated on acquiring more data to tighten the confidence bounds and increase the accuracy of the predicted in-orbit upset rate\cite{Travinsky_Rad_test_Journal_I}.\\

Here we present the results of the second data collection sets, where we continue to characterize the behavior and investigate the performance of 0.7\textsuperscript{$\prime \prime$} 1024\(\times\)768 eXtended Graphics Array (XGA) DMDs under heavy ion irradiation, using the accelerated radiation testing approach. The goal of this experiment was to increase the accuracy of the predicted heavy ion-induced single-upset rate of a DMD as a part of a Multi Object Spectrometer in an orbit located one astronomical unit (AU) away from the sun in interplanetary space, where the primary sources of radiation are sun particle fluxes and cosmic rays\cite{Shea2009}.\\

The fundamental assumption used for SEU accelerated radiation testing is that there is a sensitive region within each micromirror circuit, a sensitive volume, that can collect charge generated by the passage of a heavy ion\cite{Petersen1992}. An SEU occurs when the charge dissipated by a heavy ion in the circuit passes some critical threshold. The area of this sensitive volume is typically assumed square and can be determined from the maximum cross-section that is observed during the data collection, combined with the number of micromirrors.
The maximum cross-section observed in all the data collection sets is 1.777$\times$10$^{-7}$ cm$^2$/micromirror. Since the sensitive volume cross-section is assumed to be square, by taking the squared root of this value the dimensions of it can be found: 0.421$\times$0.421 \(\mu\)m$^2$. The thickness of the sensitive volume was chosen to be 1 \(\mu\)m, because this is the silicon depth most affected by the incident flux.\\

The spectrum of heavy ions in interplanetary space was modeled using the SPENVIS\cite{Heynderickx2004} (Space Environment Information System) software package. We assumed standard spacecraft shielding of 100 mils (2.54 mm) of spherical aluminum and used the sensitive volume dimentions described above. CREME-96\cite{Petersen1992} solar particle flux model was used to predict the worst-week solar weather on orbit. To include the Galactic Cosmic Rays (GCR) background into the model, ISO15390 with the full ion range (H to U) standard was applied. Solar activity data at solar minimum was used to ensure the worst case scenario for GCR. To estimate the  single event upset (SEU) rate, we used the integral rectangular parallelepiped (IRPP) method\cite{Petersen1983,Petersen1992,Petersen2005} with the cumulative Weibull distribution function.
%------------------------------------------------------------------------------
% Facility
%------------------------------------------------------------------------------
\subsection{The facility}
The newer heavy ion radiation testing was performed on April 17th, 2016, 8 am to 8 pm local time at the Cyclotron Institute of the Texas A\&M University (TAMU). The facility is equipped with three beam types: 15 MeV/amu (atomic mass unit), 25 MeV/amu and 40 MeV/amu. For this testing we used the 25 MeV/amu beam with two different ions – neon (Ne, linear energy transfer at normal incidence (\(\text{LET}_{\text{0}}\)) = 1.8 \(\text{MeVcm}^{\text{2}}\text{mg}^{\text{-1}} \)) and argon (Ar, \(\text{LET}_{\text{0}}\) = 5.9 \(\text{MeVcm}^{\text{2}}\text{mg}^{\text{-1}} \)). The next available ion was krypton (Kr, \(\text{LET}_{\text{0}}\) = 19.8 \(\text{MeVcm}^{\text{2}}\text{mg}^{\text{-1}} \)), but this LET was too high for our purposes in this testing, since the goal was to obtain as many data points as possible below the LET$_{\text{0}}$ of Kr \cite{Travinsky_Rad_test_Journal_I}. Therefore, to increase the LET range within this region we used two incident angles of the heavy ion beam,  \( \text{0}^{\circ} \) and \( \text{45}^{\circ} \) (measured from the normal to the surface of the device). We also used a degrader foil at both orientations to further increase the LET of Ne and Ar. The degrader foil system at TAMU Cyclotron allowed us to set the desired beam LET value without changing the beam ion or the incident angle at the target. The full resulting range of LETs for these two ions under different test configurations can be found in Table \ref{tab:test_data}.
\begin{table}[ht]
\caption{Summary of ions and their corresponding LET $\big[\text{MeVcm}^{\text{2}}\text{mg}^{\text{-1}}\big]$, angles of incidence, and beam degradation for both radiation testing data collections.} 
\label{tab:test_data}
\begin{center}       
\begin{tabular}{|l||c|c|c|c|} %% this creates two columns
%% |l|l| to left justify each column entry
%% |c|c| to center each column entry
%% use of \rule[]{}{} below opens up each row
\hline
\rule[-1ex]{0pt}{3.5ex} \multirow{2}{*}{Ion} & \multicolumn{2}{c|}{no degrader} & \multicolumn{2}{c|}{degrader in}\\
\hhline{~----}
\rule[-1ex]{0pt}{3.5ex}  & \(\text{LET}_{\text{0}}\)  & \( \text{LET}_{\text{45}} \) &  \(\text{LET}_{\text{0}}\) & \( \text{LET}_{\text{45}} \)\\
\hline\hline
\rule[-1ex]{0pt}{3.5ex}  Ne & 1.8  & 2.6 & 3.5 & 4.7 \\
\hline
\rule[-1ex]{0pt}{3.5ex}  Ar & 5.9 & 8.1 & 10 & 13   \\
\hline
\rule[-1ex]{0pt}{3.5ex}  Kr & 21.2 & 29.9 & - & -   \\
\hline
\rule[-1ex]{0pt}{3.5ex}  Xe & 41.7 & 59.2 & - & - \\
\hline
\end{tabular}
\end{center}
\end{table}
%---------------------------------------------------------------------------------------------------------------
% Test setup Radiation
%----------------------------------------------------------------------------------------------------------------
\subsection{Test setup}
\begin{figure}
\centering
\includegraphics[width=0.99 \textwidth]{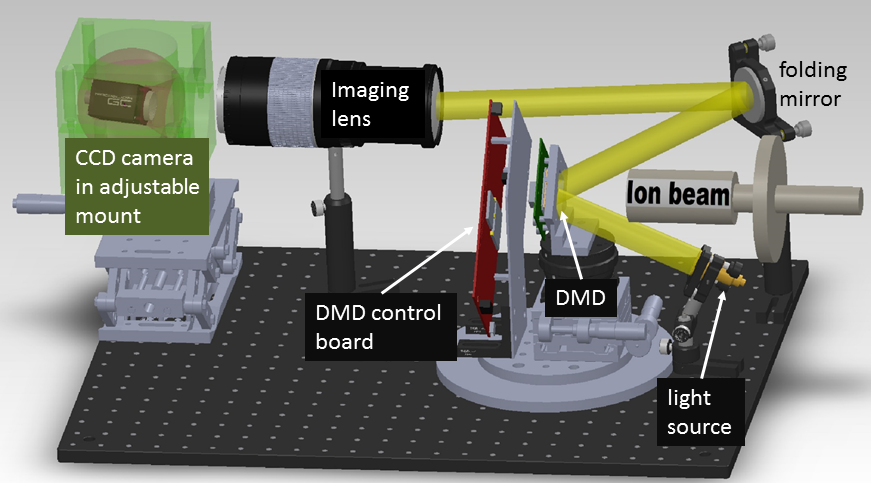}
\setlength{\abovecaptionskip}{12pt}
\caption 
{ \label{fig:TestSetup}
Test setup for the DMD radiation testing. The light from the light source illuminates the DMD surface. If the micromirrors are turned towards the folding mirror, specular reflected light reaches the imaging sensor. When micromirrors are turned the other way, all light is reflected away from the sensor.} 
\end{figure}

The approach we used to evaluate the performance of the DMD is to exercise the device under irradiation by switching all the micromirrors between the two operational states every 10 seconds. To count the number of upset micromirrors, hereafter referred to as the number of single-event upsets (SEUs), the device was illuminated by a broadband light source and an image of the device's surface was formed on an imaging detector (Figure \ref{fig:TestSetup}) with the resolution of about 1.3 detector pixel per one DMD micromirror, which is more that sufficient to detect single micromirrors. The illumination and the imaging were done off-axis, which resulted in a tilted image plane. To assist with optical alignment and focus of the system, a rotational spherical mount was designed to house the camera and permit rotation without translating the sensor relatively to the center of the image.
From the image on the detector SEUs could be detected by observing local changes in the intensity. When all the mirrors of the DMD are flipped towards the folding mirror, incoming light reaches the imaging sensor, forms a bright image of the DMD surface. A tripped mirror, due to particle-induced excessive charge or mechanical damage, may point elsewhere and its light would therefore not reach the detector. This results in a drop of the intensity at the corresponding location in the image, which can be easily detected. Turning all the micromirrors of the DMD away from the folding mirror produces a low intensity image on the sensor. In this case a stuck mirror would create a bright spot on a dark background. Initial upset count was done in real time during irradiation; the acquired images were also saved for later analysis with image processing methods.
\begin{figure}
\centering
\includegraphics[width=\textwidth]{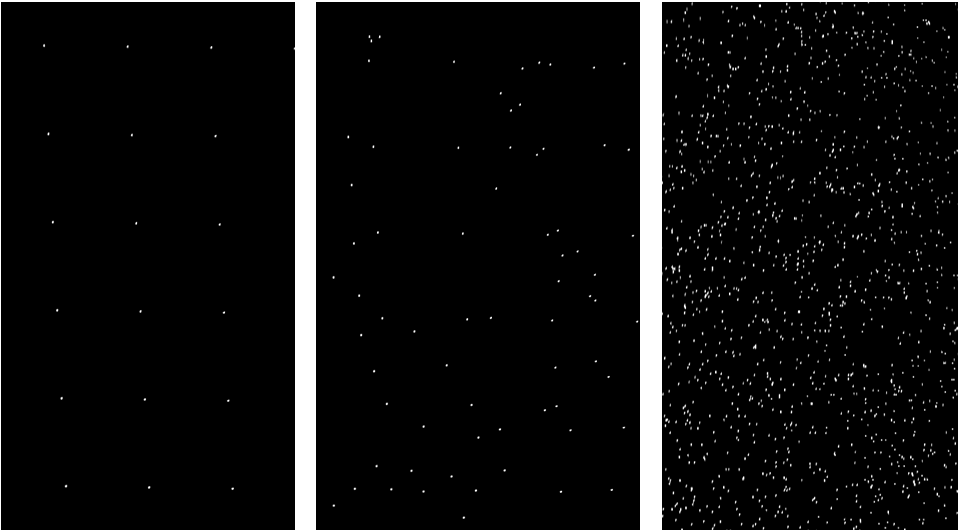}  % fig2 includes two images 
(a) \hspace{0.3\textwidth} (b) \hspace{0.3\textwidth} (c)
\setlength{\abovecaptionskip}{12pt}
\caption 
{ \label{fig:upsets}
A constant region of interest corresponding to about 310×550 micromirrors of the DMD with the best image quality with only the reference grid pattern present (a), several tens of upset mirrors present (b), several thousands of upset mirrors present (c).} 
\end{figure} 
%---------------------------------------------------------------------------------------------------------------
% Test procedure
%----------------------------------------------------------------------------------------------------------------
\subsection{Test procedure}
Commercially available DMDs (from Texas Instruments) are packaged using 3 mm thick borosilicate window, which completely attenuate the heavy ions available at TAMU Cyclotron. In order for the heavy ion flux to reach the semiconductor layer of the DMDs under test, test devices were re-windowed with imaging quality, 2 \(\mu\)m thick, low atomic number pellicles\cite{Travinsky_Rad_test_Journal_I}. The effects of such a thin and low-density pellicle on the 25 MeV ion beam are negligible and consist of a slight increase in the ion LET and a correspondingly small decrease in the mean ion range. \\
For transportation from Rochester to the TAMU cyclotron facility the devices were packed into pressurized containers\cite{Travinsky_Rad_test_Journal_I} to protect the pellicle windows from stretching due to in-flight cabin pressure changes. Two re-windowed DMDs (serial numbers 190906 and 190906) were subjected to heavy ion irradiation, while an additional re-windowed device (serial number 190907) served as a control device.\\
The experiment was conducted in the following manner: having the ion beam set on the desired ion and flux, we start the irradiation of the DMD while exercising the device by switching all the micromirrors between the \emph{on state} and the \emph{off state} (\(\pm\)12\(^{\circ}\)) every 10 seconds and acquiring an image of the device surface after every switch. Once we reach the desired fluence, we stopped the irradiation and ceased exercising the DMD. We then evaluate the number of upsets caused by the radiation with the current cyclotron parameters and decided on new settings to proceed with the next test. Having the new flux, orientation and beam degradation state set, we repeated the process described above. We referred to one single repetition of the process as one run. During the experiment we are reporting here we performed 81 runs with different combinations of ion, orientation, and beam degradation state. The summary of experiment parameters can be found in Table \ref{tab:test_data}. This combination of ions, incident angles and degradation states was specifically chosen to optimize the usage of the very limited beam time at the cyclotron and to produce the largest possible number of samples within the desired range of LET.\\
Initial evaluation of the behavior of the device was done by estimating the number of single event upsets (SEUs) in real time during the experiment, counting the number of SEUs in a predefined region of about 750\(\times\)900 micromirrors. Image acquisition timing was synchronized with micromirror switching between the active states, so if one image was acquired with micromirrors pointing toward the sensor, the next one would be with micromirrors pointing away. In the first case we were looking for dark spots on bright background, and in the later one - for bright spots on dark background (Figure \ref{fig:upsets}). The number of upsets was displayed for initial feedback, so as to be able to make effective decisions about the quantitative parameters for the following runs of the experiment. All images were saved with the corresponding meta data for later evaluation.
%----------------------------------------------------------------------------
% Data analysis and results
%----------------------------------------------------------------------------
\subsection{Data analysis and results}
\label{subsec:RadTestResults}
\begin{figure}
\centering
\includegraphics[width=0.99 \textwidth]{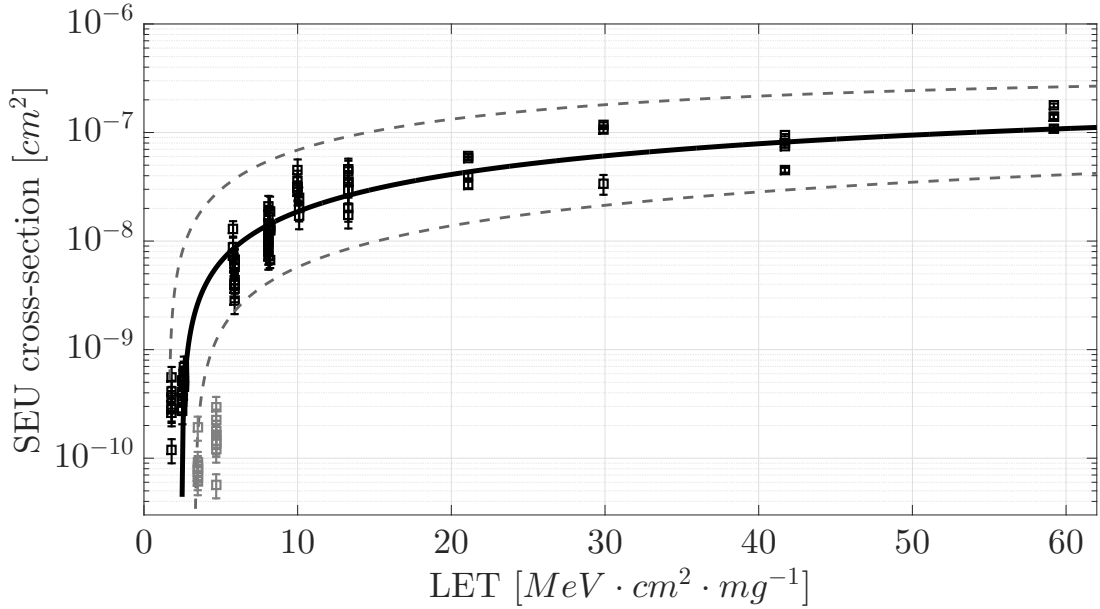}
\setlength{\abovecaptionskip}{12pt}
\caption 
{ \label{fig:RadTestResults}
Measured SEU cross-section as a function of LET of the incident ions. The Weibull cumulative distribution fit with 95\% confidence level bounds is presented as the solid line and the dotted lines respectively. The outliers marked in gray are the runs obtained using the Ne beam with a degrader. These measurements did not provide reliable data due to extremely high dispersion of the low energy Ne ions by the degrader, and were therefore high mismatch between the theoretical and the actual LET values. These runs were not included in the fitting.} 
\end{figure}

In order to obtain the parameters required for in-orbit single event upset rate estimation, the experimental data has to be reduced to a plot of SEU cross-section as a function of LET\cite{JESD89A} (Figure \ref{fig:RadTestResults}). Cross-section was calculated by normalizing the number of upset mirrors by the overall number of mirrors in the region of interest and by the number of heavy ions incident on the device per unit area (fluence). The experimental parameters - fluence, flux, effective LET, measurement uncertainty, beam uniformity, and active beam times are logged and provided as text files by the TAMU facility. After all the experimental data was reduced to a single SEU cross-section against LET plot (Figure \ref{fig:RadTestResults}), fitting to a Weibull distribution was performed to obtain the parameters for the SEU rate estimation with the SPENVIS software package\cite{Heynderickx2004}, using the parameters described in Section \ref{subsec:RadtestBackground}.\\
The most general form of the Weibull distribution is the 4-parameter one:
\begin{equation}
\sigma(x) = \sigma_{\text{lim}} \Bigg[1-\text{e}^{-\Big(\frac{x-x_0}{W} \Big)^s} \Bigg]
\end{equation}
where:
\begin{conditions}
 \sigma  & is the SEU cross-section [cm$^2$]\\
 \sigma_{lim}  & is the plateau SEU cross-section [cm$^2$]\\
 x_0   & is the minimal amount of energy per unit length required to cause an upset [MeVcm$^2$ mg$^{-1}$]\\
W   &    is the width of the range of LET, where the SEUs are observable $\sigma(x_0+W)=\sigma_{lim}$ [MeVcm$^2$ mg$^{-1}$]\\
s   &    is the slope parameter that determines the shape of the Weibull distribution. 
\end{conditions}
Although the main cause of failure for electrostatic MEMS devices at high radiation doses is the accumulation of charge in dielectric
layers\cite{Shea2011}, in this case of irradiation of DMDs with heavy ions, the total dose effects are negligible in comparison with transient single event effects (SEEs). The only type of SEEs we observed in our experiments is single event upsets - SEUs, which are temporary and can be cleared by uploading a new pattern and sending a global reset pulse to the DMD. We believe that the cause of the SEUs were passing heavy ions that caused accumulation of charges in the DMD CMOS layer. These charges probably tripped the state of DMD memory cells into the opposite from the one the cell was originally set while uploading a pattern into the DMD memory. Therefore, at a later point, when the upload is complete and a global reset is applied, this tripped cell sets the corresponding micromirror into a false position operating position (e.g. into the \emph{on state} instead of the \emph{off state}). The number of SEUs in this case only depends on the fluence and LET, and therefore the SEU rate is constant, which means that the Weibull distribution becomes exponential (s=1). 

Fitting to the Weibull distribution at the 95\% confidence level was performed in MATLAB using the Nonlinear Least Squares algorithm. The slope parameter was set to one to ensure an exponential distribution shape. The outliers marked on the graph (Figure 4) were excluded. The model coefficients obtained from the fitting to the Weibull distribution and the 95\% confidence bounds are: 
\begin{equation}
\begin{split}
& \sigma_{\text{lim}} = \text{2.24} \times \text{10}^{\text{-7}} (\text{1.21} \times \text{10}^{\text{-7}}, \text{3.27} \times \text{10}^{\text{-7}}) \text{ cm}^{\text{2}} \\
& \text{W = 86.59  (35.32, 137.8)} \text{ MeVcm}^2 \text{mg}^{\text{-1}} \\
& \text{x}_{\text{0}} = \text{2.456  (1.62, 3.31)} \text{ MeVcm}^2\text{mg}^{\text{-1}} \\
& \text{R}^\text{2} = \text{0.96}.
\end{split}
\end{equation}  
Using the Weibull distribution coefficients from the fitted model, we calculated the upset rate on orbit for the worst week case using the SPENVIS software package\cite{Heynderickx2004}. We assumed one-year duration mission in interplanetary space at a distance of one AU from the Sun. We used CREME-96\cite{Petersen1992} solar particle flux model and the full (H to U) ion range for the worst week scenario. The SEU rate from direct ionization resulted in 7.0907$\times$10$^{-6}$ micromirrors$^{-1}$day$^{-1}$. Considering that the tested type of device was a micromirror array of 1024$\times$768 mirrors, the overall predicted on orbit non-destructive micromirror upset rate is 5.57 per day, or less than one micromirror every four hours, in the worst week scenario. Using the 95\% confidence bounds obtained from the fitted model we calculate lower and upper values for the predicted worst-case in-orbit SEU rate to be [3.42, 11.7] micromirrors in 24 hours.
%%%%%%%%%%%%%%%%%%%%%%%%%%%%%%%%%%%%%%%%%%%%%%%%%%%%%%%%%%%%%%%%%%%%%%%%%%%%%
% Vibration and Shock testing
%%%%%%%%%%%%%%%%%%%%%%%%%%%%%%%%%%%%%%%%%%%%%%%%%%%%%%%%%%%%%%%%%%%%%%%%%%%%%
\section{Vibration and Shock testing}
\label{sec:vibration}
Along with radiation environment, another important consideration for space-based instruments is operability under the vibration and mechanical shock that is typically associated with launch. Space-based DMD-based MOSs are not an exception and therefore the behavior of DMD under vibration needed to be characterized, as specified by the NASA General Environmental Verification Standard (GEVS)\cite{GEVS}.\\
DMDs are very tolerant to vibration at low frequencies because the moving parts inside these devices are on the order of tens of microns in size and the range of mirror movement is in single microns $\mu$m for the 0.7\textsuperscript{$\prime \prime$} XGA devices. Therefore the lowest resonance frequency of the micromirrors inside DMD is on the order of several hundreds of kilohertz \cite{Douglass2003}. As a matter of fact, the packaging of DMDs is more of a concern under vibration load than the micromirrors themselves \cite{Robberto2009}. Moreover, vibration tests on DMD have been previously performed by TI\textsuperscript{TM}; they tested DMDs in the range of 20-2000 Hz at up to 20 g, and mechanical shock tests at up to 1500 G with no failures \cite{Douglass2003}. Nevertheless, the vibration testing conducted by TI\textsuperscript{TM} was done to assess the reliability of DMD for ground-based applications, and not according to NASA standards. The experiments reported here were performed according to the NASA GEVS\cite{GEVS} for space-based applications. The purpose of our test was to assess the affects of vibration on the DMD micromirrors themselves and on the hermeticity of the packaging of the devices.
%----------------------------------------------------------------------------
% Vibration and Shock testing-->Window replacement and leak testing
%----------------------------------------------------------------------------
\subsection{Window replacement}
\label{subsec:WindowReplacement}
\begin{figure}
\centering
\includegraphics[width=\textwidth]{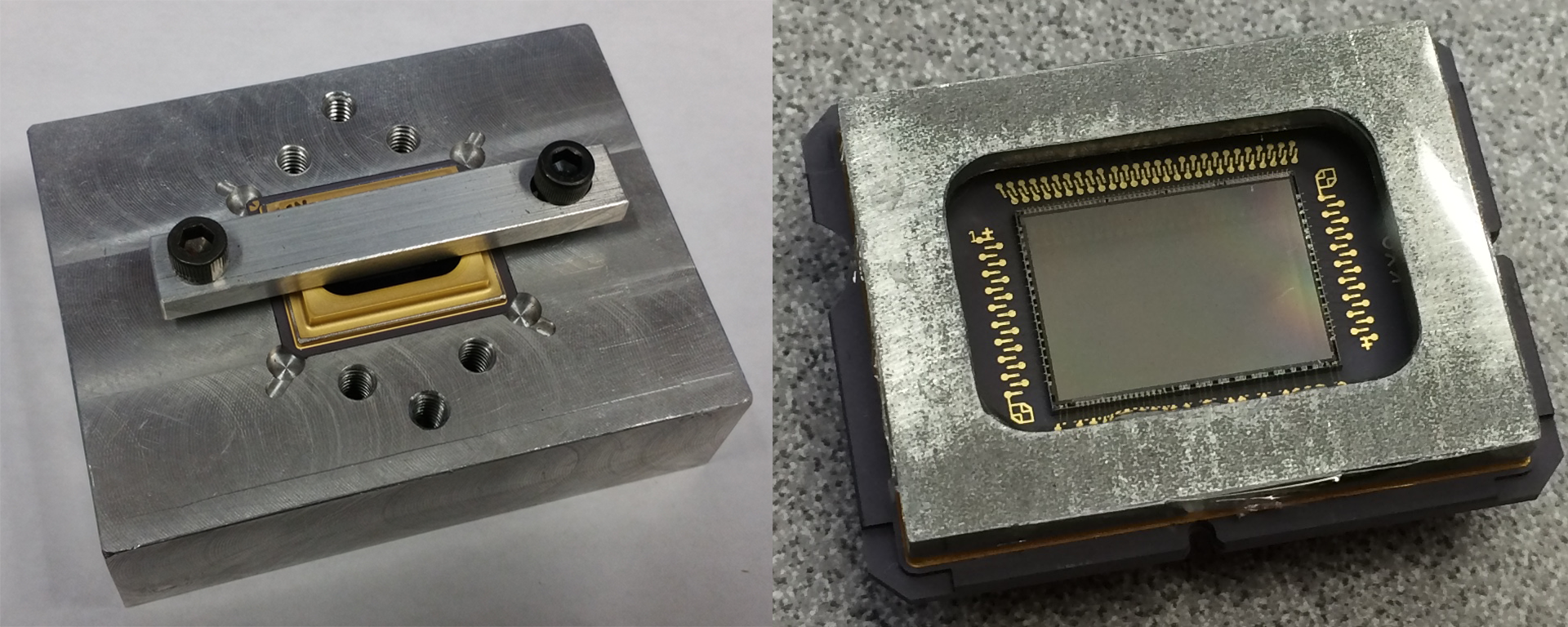} 
\setlength{\abovecaptionskip}{12pt}
\caption 
{ \label{fig:rewindowedDMDs}
(\textit{left}) A mechanical fixture is used to hold the DMD, while a CNC mill removes the weld which attaches the window frame to the DMD package. (\textit{right}) A stainless steel frame is bonded to the DMD package using an indium seal and low out-gassing epoxy. The stainless steel frame can support a wide range of window materials, such as thin cellulose film (shown here), as well as fused silica, sapphire and MgF$_2$.} 
\end{figure}

DMDs built by TI\textsuperscript{TM} are protected from damage due to dust, moisture, and direct contact by a hermetically sealed package with a borosilicate window. However, the  window is opaque to ultra-violet (UV) or infra-red (IR) radiation, which limits their utility in astronomy. To overcome this limitation, we retro-fitted commercially available DMDs with windows made from magnesium fluoride (MgF$_2$), ultra-violet (UV)-grade heat exchange method (HEM)-grown sapphire, fused silica, and even 2 $\mu$m thick cellulose triacetate film \cite{Travinsky_Rad_test_Journal_I}. We previously described the re-windowing process\cite{Travinsky_Rad_test_Journal_I} in great detail. It consists of the following steps:
\begin{itemize}
\item[Step 1:] The DMD is placed in a mechanical fixture, which clamps the window frame to the ceramic DMD package
(Figure \ref{fig:rewindowedDMDs}, \textit{left} ).
\item[Step 2:] Once the seam has been mechanically severed, the DMD and fixture assembly is transferred to a class
1000 clean room where the clamp is removed and the window is separated from the package by a light
shearing force.
\item[Step 3:] The window material of choice is attached to a stainless steel frame using a NASA-rated low-outgassing
epoxy (EPO-TEK 353ND) which has been fully cured in a prior processing step.
\item[Step 4:] The frame is then mounted on the DMD assembly with a crushed indium wire gasket surrounding the DMD chip cavity (Figure \ref{fig:rewindowedDMDs}, \textit{right}). This creates an air-tight seal, which keeps out dust and moisture.
\item[Step 5:] Optionally, the bond between the stainless steel frame and the DMD package can be mechanically reinforced by a bead of low-outgassing epoxy applied to the perimeter seam.
\end{itemize}
%----------------------------------------------------------------------------
% Vibration and Shock testing-->Window replacement and leak testing-->Evaluating seal integrity
%----------------------------------------------------------------------------
\subsection{Leak testing}
\label{subsec:VibTest:LeakTesting}
\begin{figure}[htb!]
\centering
\includegraphics[width=\textwidth]{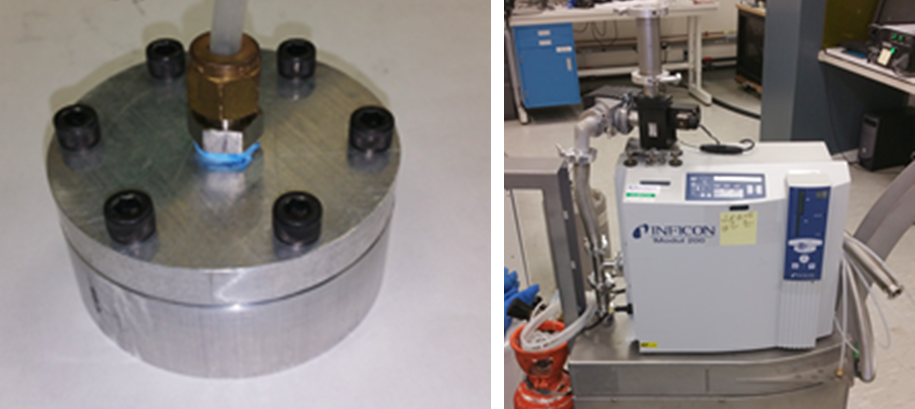}  % fig6 includes two images 
\setlength{\abovecaptionskip}{12pt}
\caption 
{ \label{fig:leakTestChamber}
The re-windowed DMDs were tested for leaks to determine the integrity of the new seal: (\textit{left}) The DMDs were placed in an aluminum ``bomb" chamber with a Viton o-ring seal and pressurized in a pure helium atmosphere; (\textit{right}) An Inficon helium mass spectrometer was used to measure trace amounts of helium diffusing out of the DMDs. The leak
detector is equipped with an internal NIST-traceable calibrated leak to conveniently verify the system leak rate calibration prior to measurement.} 
\end{figure}

Measurement of the hermetic seal quality of optoelectronic packages was performed using the MIL-STD-833 test method 1014. We utilized the test condition A$_\text{2}$ ``flexible" method which describes a fine leak test using a helium trace gas. In this method, a component under test is placed in an ampoule called a ``bomb" (Figure \ref{fig:leakTestChamber}, \textit{left}) and pressurized with pure helium gas for a specific period. Once the ``bombing" process is finished, the component is removed to a helium mass spectrometer and placed under vacuum in a small vessel (Figure \ref{fig:leakTestChamber}, \textit{right}). After a specified time in the vacuum, the partial pressure of helium detectable in the vessel is recorded and compared to the maximum allowable leak rate of the component as calculated using the Howell-Mann equation:
\begin{equation}
R_1 = \dfrac{LP_E}{P_0} \Bigg( \dfrac{M_A}{M} \Bigg)^{1/2} \Bigg( 1-e^{- \dfrac{Lt_1}{VP_0} \Big( \dfrac{M_A}{M} \Big)^{1/2}} \Bigg)e^{-\dfrac{Lt_2}{VP_0} \Big( \dfrac{M_A}{M} \Big)^{1/2}},
\end{equation}
where:
\begin{conditions}
R_1  & measured leak rate of helium in the mass spectrometer after the dwell time t$_2$\\
L & the maximum allowable equivalent standard leak rate limit in atm-cm$^3$/s air\\
P_E & the pressure of helium exposure to the helium ``bomb" in atmospheres (absolute)\\
P_0 & the atmospheric pressure in atmospheres (absolute)\\
M & molecular weight of helium\\
t_1 & helium bomb dwell time in seconds\\
t_2 & dwell time in vacuum between release of pressure and leak detection in seconds\\
V & internal free volume of the device package cavity in cm$^3$.
\end{conditions}
We leak-checked devices before and after the vibration testing, while the commercial TI\textsuperscript{TM} packages were used for reference.\\

The first ``bombing" step is performed using a simple aluminum chamber that can be pressurized from a helium bottle through a regulator. The chamber is filled with helium and the pressure is released to purge the atmosphere from the chamber over five cycles. We have chosen P$_\text{E}$=60 pounds per square inch gage (PSIG) as the bombing pressure, for a period of t$_{\text{1}}$=3 hours. Any small leaks will pressurize the component interior with some partial pressure of helium. \\

The second step is performed in a stainless steel chamber mounted on a mass spectrometer helium leak detector. The part is inserted into the chamber and placed under vacuum for a dwell time t$_2$ = 1 hour. After time t$_2$, the helium leak rate is recorded and converted to a leak rate in air.\\

The internal cavity of a standard DMD package is about 23$\times$33$\times$2.5 mm = 1.8 cm$^3$\cite{DMDdatasheet}.The volume of the active area of the DMD can be calculated as 13$\times$11 mm$\times$1  mm\cite{DMDdatasheet} = 0.143 cm$^3$. Subtracting the latter from the former leaves the cavity volume of ~1.65 cm$^3$ for the standard TI\textsuperscript{\texttrademark}-packaged DMDs. The re-windowed devices have slightly larger cavity of 2.83 cm$^3$ because the steel frame (Section \ref{subsec:WindowReplacement}, Step 3) that adds more volume.\\

According to MIL-STD-883\cite{MILSTD883E} the failure criterion L for packages with internal volume larger than 0.5 cm$^3$ is 1 $\times$10$^{-7}$ atm-cm$^3$/s of air. Thus, the reject limit R$_1$ of devices under these conditions is 1.4$\times$10$^{-7}$ atm-cm$^3$/s of helium, which corresponds to 5.2$\times$10$^{-6}$ atm-cm$^3$/s of air. Parts as-received from Texas Instruments Inc. will typically exhibit very low leak rates in the 10$^{-10}$ atm-cm$^3$/s He range, corresponding to a very high integrity welded hermetic package. Re-windowed DMD devices using epoxy sealing materials typically exhibit leak rates in the range 10$^{-7}$ atm-cm$^3$/s of helium indicating a near-hermetic seal. The leak rates for the devices before the vibration testing are shown in Table \ref{tab:vibration_leak_test}.
\begin{table}[ht]
\caption{Leak rates for DMDs before the vibration testing.} 
\label{tab:vibration_leak_test}
\begin{center}       
\begin{tabular}{|l|l|l|l|} %% this creates two columns
%% |l|l| to left justify each column entry
%% |c|c| to center each column entry
%% use of \rule[]{}{} below opens up each row
\hline
\rule[-1ex]{0pt}{3.5ex}  Serial number & Window  material  & He leak rate (atm-cm$^3$/s) &  Notes \\
\hline\hline
\rule[-1ex]{0pt}{3.5ex}  190806 & fused silica &  6.00$\times$10$^{-7}$ &  stainless frame  \\
\hline
\rule[-1ex]{0pt}{3.5ex}  190805 & fused silica &  6.00$\times$10$^{-7}$ &  stainless frame    \\
\hline
\rule[-1ex]{0pt}{3.5ex}  191306 & sapphire &  1.30$\times$10$^{-7}$ &     \\
\hline
\rule[-1ex]{0pt}{3.5ex}  191207 & magnesium fluoride &  1.00$\times$10$^{-7}$ &    \\
\hline
\rule[-1ex]{0pt}{3.5ex}  110306 & borosilicate &  6.40$\times$10$^{-10}$ &    \\
\hline
\end{tabular}
\end{center}
\end{table}
%----------------------------------------------------------------------------
% Vibration and Shock testing-->Testing methodology and setup
%----------------------------------------------------------------------------
\subsection{Testing methodology and setup}
\begin{figure}
\centering
\includegraphics[width=\textwidth]{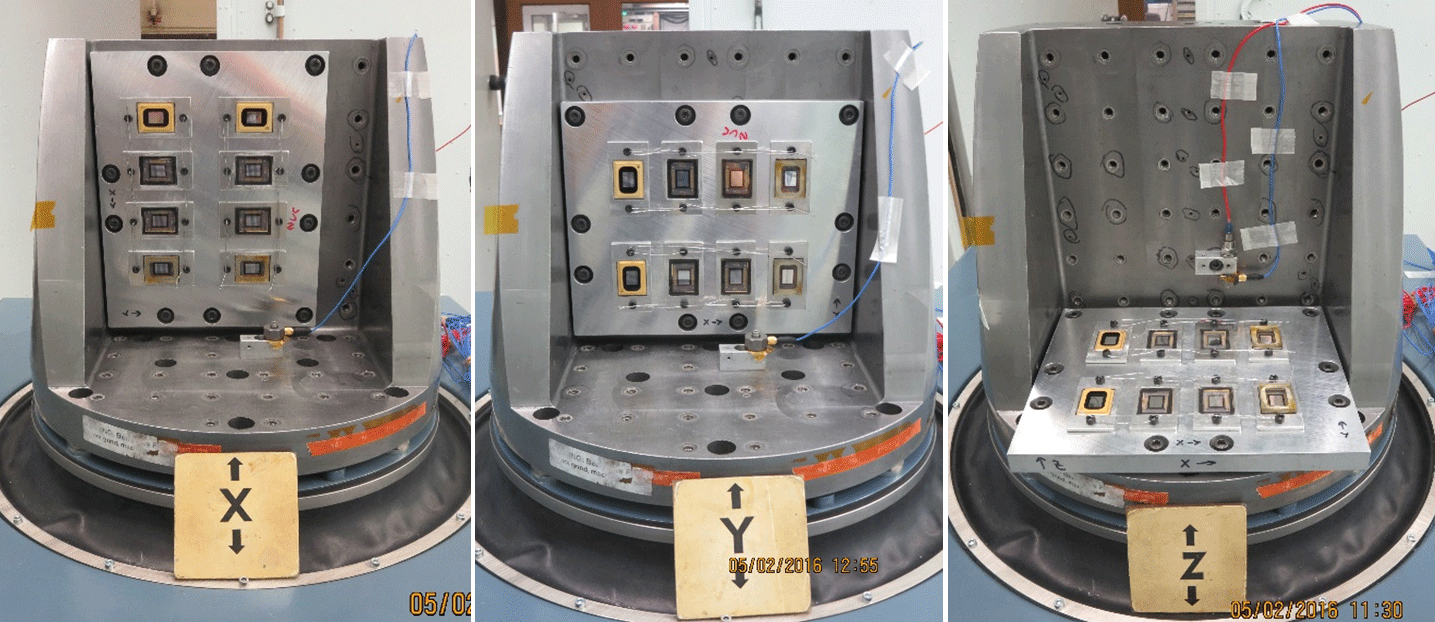} 
\setlength{\abovecaptionskip}{12pt}
\caption 
{ \label{fig:fixturePlates}
The plate which held up to 8 DMDs at a time was mounted to the AVCO fixture on the T2000-2 shaker. The plate orientation was changed to perform tests along the X, Y and Z axes of the DMDs, because the T2000-2 shaker only has 1 axis of motion. The accelerometer placement was also changed to probe the appropriate axis.} 
\end{figure}

Our objective was to investigate the capability of the DMDs to survive the rigors of potential vibroacoustic launch environments via the NASA General Environmental Verification Standard (GEVS)\cite{GEVS} sine burst, random, and mechanical shock testing. The vibration and shock tests were carried out May 2nd - 3rd, 2016 at the Sierra Lobo Vibration Facility at NASA's Goddard Space Flight Center. The tests were performed on the T2000-2 shaker, using a custom mounting plate and the AVCO fixture (Figure \ref{fig:fixturePlates}). We tested the DMDs in the not powered state as well as while powered on, and operational. We tested a total of 13 DMDs, with a variety of window materials; some DMDs were inoperable mechanical samples which were only used for testing re-windowed package integrity.
\begin{table}[ht]
\caption{Summary of DMDs subjected to vibration and shock testing.} 
\label{tab:vibration_devices_summary}
\begin{center}       
\begin{tabular}{|l|l|l|l|} %% this creates two columns
%% |l|l| to left justify each column entry
%% |c|c| to center each column entry
%% use of \rule[]{}{} below opens up each row
\hline
\rule[-1ex]{0pt}{3.5ex}  Serial number & Window  material  & Test load (dB) &  Notes \\
\hline\hline
\rule[-1ex]{0pt}{3.5ex}  190806 & fused silica &  -12, -6, 0 &  stainless frame  \\
\hline
\rule[-1ex]{0pt}{3.5ex}  190805 & fused silica &  -12, 0 &  stainless frame    \\
\hline
\rule[-1ex]{0pt}{3.5ex}  191007 & fused silica &   0 &      \\
\hline
\rule[-1ex]{0pt}{3.5ex}  191006 & fused silica &   0 &      \\
\hline
\rule[-1ex]{0pt}{3.5ex}  191207 & magnesium fluoride &  -12, -6, 0 &    \\
\hline
\rule[-1ex]{0pt}{3.5ex}  240207 & magnesium fluoride &  -12, 0 &    \\
\hline
\rule[-1ex]{0pt}{3.5ex}  191205 & sapphire &  -12, -6,  0 &    \\
\hline
\rule[-1ex]{0pt}{3.5ex}  191306 & sapphire &  -12,  0 &    \\
\hline
\rule[-1ex]{0pt}{3.5ex}  191005 & sapphire &   0 &   New AR coat \\
\hline
\rule[-1ex]{0pt}{3.5ex}  191105 & sapphire &   0 &   New AR coat \\
\hline
\rule[-1ex]{0pt}{3.5ex}  110306 & borosilicate &  -12, -6, 0 & standard package, mechanical sample   \\
\hline
\rule[-1ex]{0pt}{3.5ex}  250305 & borosilicate &  -12, 0 & standard package   \\
\hline
\rule[-1ex]{0pt}{3.5ex}  250408 & borosilicate &  0 & standard package   \\
\hline
\end{tabular}
\end{center}
\end{table}

%----------------------------------------------------------------------------
% Vibration and Shock testing-->Testing methodology and setup-->test facility
%----------------------------------------------------------------------------
\subsubsection{Test facility}\textbf{}
The UD T2000-2 shaker, rotated to the vertical configuration and the AVCO Fixture attached, provided the X, Y, and Z axes of vibration excitation (Figure \ref{fig:fixturePlates}). The T2000-2 is mounted on an isolation block located in Building 7 of the NASA Goddard Space Flight Center, Test Cell 026. Two tri-axial PCB Model 354T accelerometers mounted in the axis of excitation provided the control system acceleration data for closed loop shaker control for vibration testing. Two orthogonal axis responses, from one PCB 354T accelerometer provided cross axis acceleration monitoring. A shock accelerometer, Endevco Model 2225, mounted on the Vibration Facility Fixture Plate in the direction of vibration excitation provided control system acceleration data for open loop shaker control for mechanical shock vibration testing only. The mounting orientations of the DMD plate and the accelerometers are shown in Figure \ref{fig:fixturePlates}. A DSPCon DataFlex-1000 recorder in the Vibration Facility Control Room recorded the control channels, response instrumentation brought into the controller, and the backup monitor. An Unholtz Dickie Model D33 provided signal conditioning of the control and cross talk accelerometers. A Trig-Tek Model 620B provided signal conditioning for the backup monitor accelerometer. An M+P VCP9000 Vibration Control System (VCS) processed the conditioned control acceleration data for update of the exciter drive signal based on maximum levels during the vibration tests.
%----------------------------------------------------------------------------
% Vibration and Shock testing-->Testing methodology and setup-->random vibration testing
%----------------------------------------------------------------------------
\subsubsection{Random Vibration Testing}
\begin{figure}
\centering
\includegraphics[width=\textwidth]{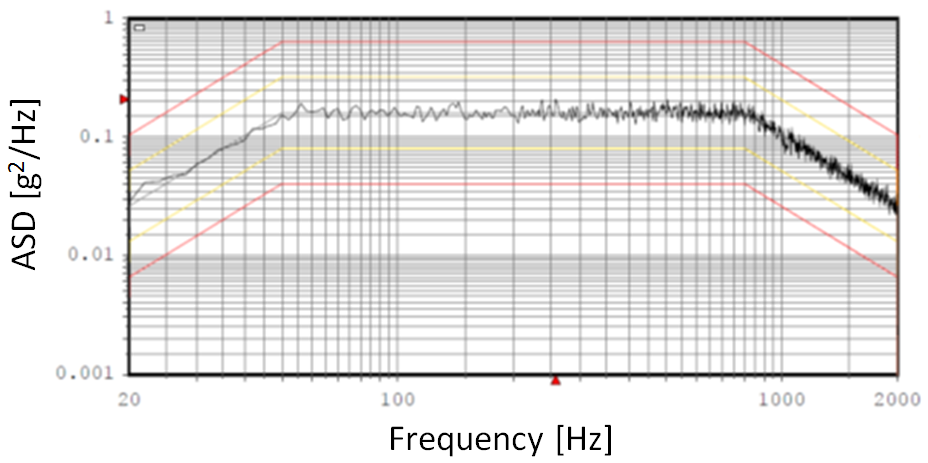} 
\setlength{\abovecaptionskip}{12pt}
\caption 
{ \label{fig:randomVibrationProfile}
An example acceleration power density spectrum for the random vibration test at 0 dB. The test duration was
60 seconds for each axis, at -18 dB, -12 dB, -6 dB and 0 dB levels.} 
\end{figure}

The random vibration test delivered 14.14 $\text{g}_{\text{RMS}}$ across a frequency range of 20 Hz - 2 kHz for 60 seconds to each axis. Some specific values of the acceleration power density are given in Table \ref{tab:random_vibration_summary} and an example acceleration density spectrum is shown in Figure \ref{fig:randomVibrationProfile}. During random vibration testing, The M+P VCS calculated an updated drive signal based on the PSD derived from the maximum g$^2$/Hz of PSDs calculated for each of the two feedback control accelerometer measurements at each frequency (spectral line). The M+P VCS compared the calculated PSD to the random vibration test specification, calculated the error, and then calculated a new PSD that accounted for the error and would meet the random vibration test specification. The M+P VCS calculated an updated drive signal from the new PSD.
\begin{table}[ht]
\caption{Random vibration acceleration spectral density (ASD) at 0 dB for several frequencies over the full frequency range, for the X, Y, Z-axis tests.} 
\label{tab:random_vibration_summary}
\begin{center}       
\begin{tabular}{|l|l|} %% this creates two columns
%% |l|l| to left justify each column entry
%% |c|c| to center each column entry
%% use of \rule[]{}{} below opens up each row
\hline
\rule[-1ex]{0pt}{3.5ex}  Frequency [Hz] & ASD [g$^2$/Hz]   \\
\hline\hline
\rule[-1ex]{0pt}{3.5ex}  20 & 0.026  \\
\hline
\rule[-1ex]{0pt}{3.5ex}  50 & 0.16   \\
\hline
\rule[-1ex]{0pt}{3.5ex}  800 & 0.16   \\
\hline
\rule[-1ex]{0pt}{3.5ex}  2000 & 0.026  \\
\hline
\rule[-1ex]{0pt}{3.5ex}  over all & 14.14 $\text{g}_{\text{rms}}$    \\
\hline
\end{tabular}
\end{center}
\end{table}
%----------------------------------------------------------------------------
% Vibration and Shock testing-->Testing methodology and setup-->Sine Burst Vibration Testing
%----------------------------------------------------------------------------
\subsubsection{Sine Burst Vibration Testing}
\begin{figure}
\centering
\includegraphics[width=\textwidth]{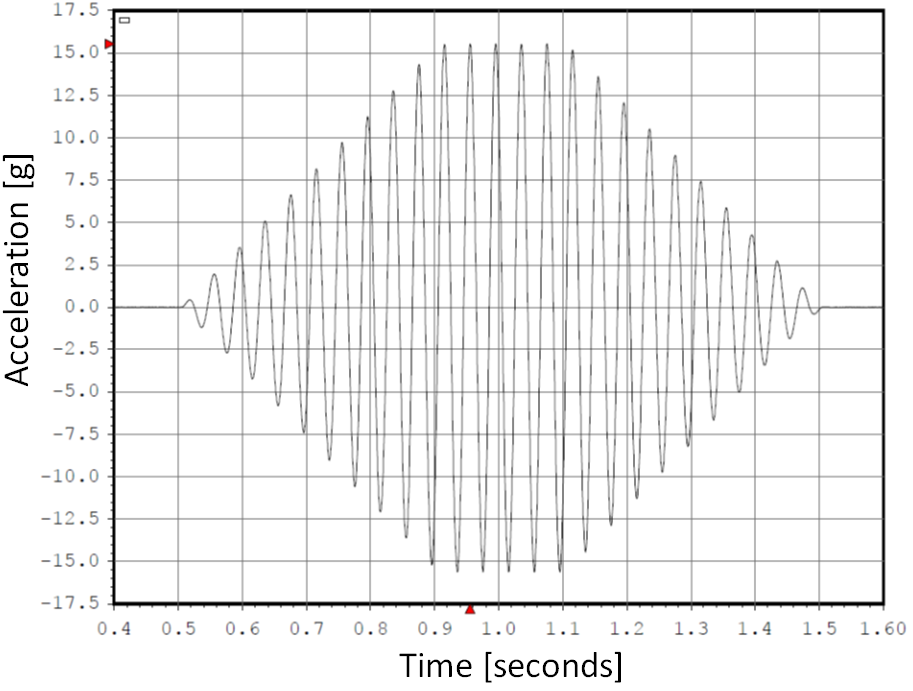} 
\setlength{\abovecaptionskip}{12pt}
\caption 
{ \label{fig:SineBurstProfile}
An example sine burst pulse delivered to each axis of the DMDs at 0 dB. Each axis was tested for 5 cycles at
-18 dB, -12 dB, -6 dB and 0 dB levels.} 
\end{figure}

The sine burst test delivered a peak acceleration of 15 g at 25 Hz to each axis of the device for 5 cycles (Figure \ref{fig:SineBurstProfile}). During sine burst vibration testing, the M+P Vibration Control System calculated the drive signal by one of two methods. For the 18 dB and 12 dB sine burst levels the M+P VCS calculated a new drive from the average of five previous sine burst pulse responses obtained from control accelerometer No. 1. For the 6 dB and 0 dB levels the M+P VCS scaled to the level from the set of averages obtained from the 12 dB level. This vibration control strategy allowed for calculation and update of the M+P VCS drive at low levels and only one sine pulse at the 6 dB and 0 dB levels.
%----------------------------------------------------------------------------
% Vibration and Shock testing-->Testing methodology and setup-->Mechanical Shock Testing
%----------------------------------------------------------------------------
\subsubsection{Mechanical Shock Testing}
\begin{figure}[tb]
\centering
\includegraphics[width=\textwidth]{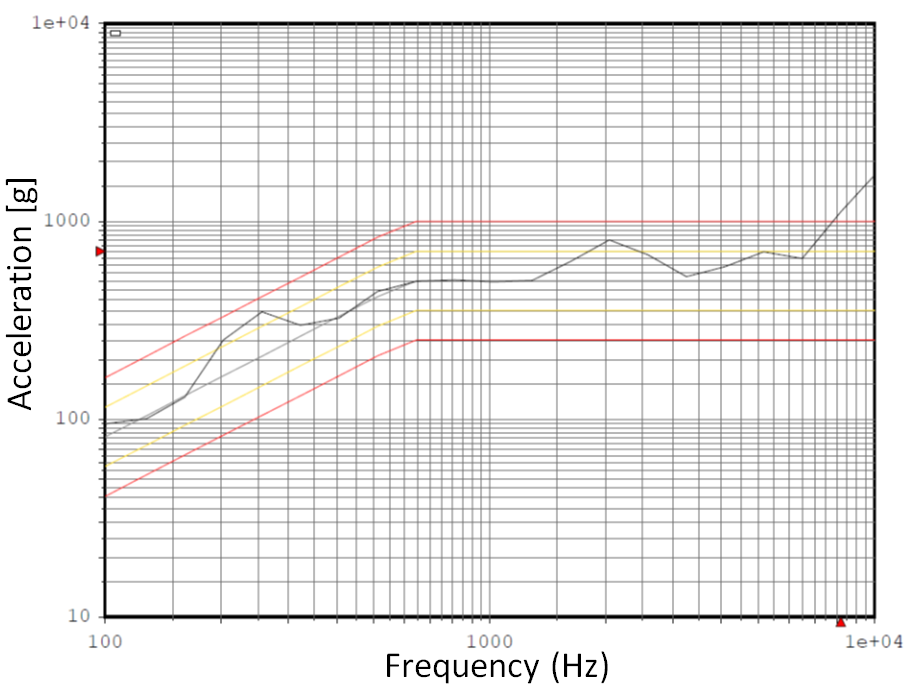} 
\setlength{\abovecaptionskip}{12pt}
\caption 
{ \label{fig:MechanicalShockProfile}
An example shock response spectrum for the mechanical shock test at 0 dB.} 
\end{figure}

The mechanical shock tests generated a shock response spectrum (SRS) that resulted in accelerations of 80 - 1000 g across a frequency range of 100 Hz - 10 kHz, for each axis. Table \ref{tab:MechanicalShockSummary} summarizes the mechanical shock test parameters and Figure \ref{fig:MechanicalShockProfile} shows an example shock response spectrum for one of the tests. During shock testing, the M+P Vibration Control System calculated the drive signal for the 18 dB level from the self-check. For the remainder of the levels 12 dB through 0 dB levels the M+P VCS scaled to the level from the set of level from the 18 dB level. This shock control strategy allowed for calculation and update of the M+P VCS drive at low levels and only one sine pulse at the 18 dB through -1 dB and 2 shocks at 0 dB levels.
\begin{table}[ht]
\caption{Mechanical shock test: shock response spectrum levels for all three axes.} 
\label{tab:MechanicalShockSummary}
\begin{center}       
\begin{tabular}{|l|l|} %% this creates two columns
%% |l|l| to left justify each column entry
%% |c|c| to center each column entry
%% use of \rule[]{}{} below opens up each row
\hline
\rule[-1ex]{0pt}{3.5ex}  Frequency [Hz] & SRS [g]   \\
\hline\hline
\rule[-1ex]{0pt}{3.5ex}  100 & 81.4  \\
\hline
\rule[-1ex]{0pt}{3.5ex}  614.5 & 500   \\
\hline
\rule[-1ex]{0pt}{3.5ex}  10000 & 500   \\
\hline
\end{tabular}
\end{center}
\end{table}

%----------------------------------------------------------------------------
% Vibration and Shock testing-->Testing methodology and setup-->Vibration Testing of Powered-on DMDs
%----------------------------------------------------------------------------
\subsubsection{Vibration Testing of Powered-on DMDs}
\begin{figure}
\begin{center}
\begin{tabular}{c}
\includegraphics[width=\textwidth]{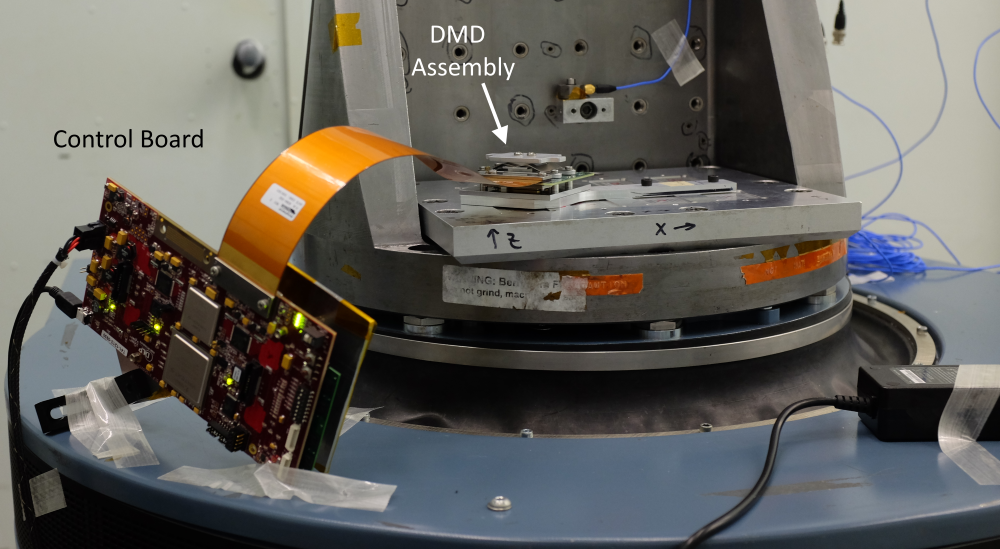} 
\\
\end{tabular}
\end{center}
\caption 
{ \label{fig:Powered-onVibrationTesting}
Vibration testing of DMDs in powered-on state. Two configurations: while displaying a static pattern and while rapidly cycling through several patterns. The entire DMD assembly was mounted on the custom plate, with the control board nearby. The DMD was only tested in the Z-axis in the powered-on state, using the same procedures and levels as for the rest of tests.} 
\end{figure}

\begin{figure}
\centering
\includegraphics[width=\textwidth]{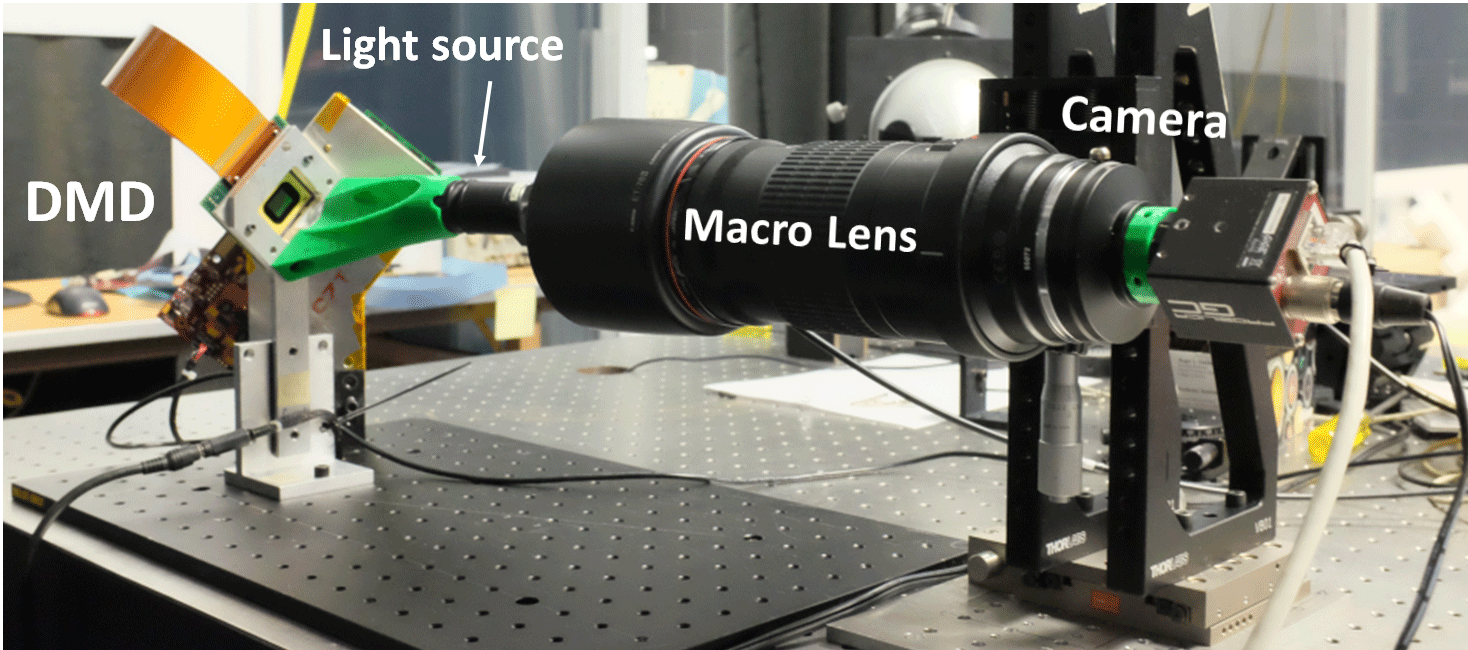} 
\setlength{\abovecaptionskip}{12pt}
\caption 
{ \label{fig:VibrationTestingImagingSetup}
The DMDs were inspected with a 2:1 macro zoom lens to detect any micromirrors that were tripped as a result of the tests.} 
\end{figure}

In addition to testing powered off DMDs, we tested the DMD in two operating modes: holding a static pattern and rapidly switching between pre-loaded patterns. In the first case, a static reference pattern was displayed on the DMD and images of the DMD were acquired to record the pre-test condition of the device. Next, without powering down the DMD or changing the pattern, the DMD assembly was mounted on the custom mounting plate in the ``face down" Z-axis configuration (Figure \ref{fig:Powered-onVibrationTesting}). We performed random vibration tests, sine burst and mechanical shock tests at 0 dB on the powered-on DMD using the procedures described in the above sections. Due to time constraints, we chose to only test the Z-axis, because we expect this to be the orientation most susceptible to damage. After the tests, we transferred the DMD assembly, without switching the device off or uploading a new pattern, to our optical inspection setup (Figure \ref{fig:VibrationTestingImagingSetup}) and inspected the DMD for pixels that may have changed their state (i.e., the direction of the tilt of the micromirrors). We did not detect any pixels that changed their state or became inoperable as a result of the tests.
%----------------------------------------------------------------------------
% Vibration and Shock testing-->Testing methodology and setup-->Detection of tripped mirrors
%----------------------------------------------------------------------------
\subsubsection{Detection of tripped mirrors}
\begin{figure}
\centering
\includegraphics[width=\textwidth]{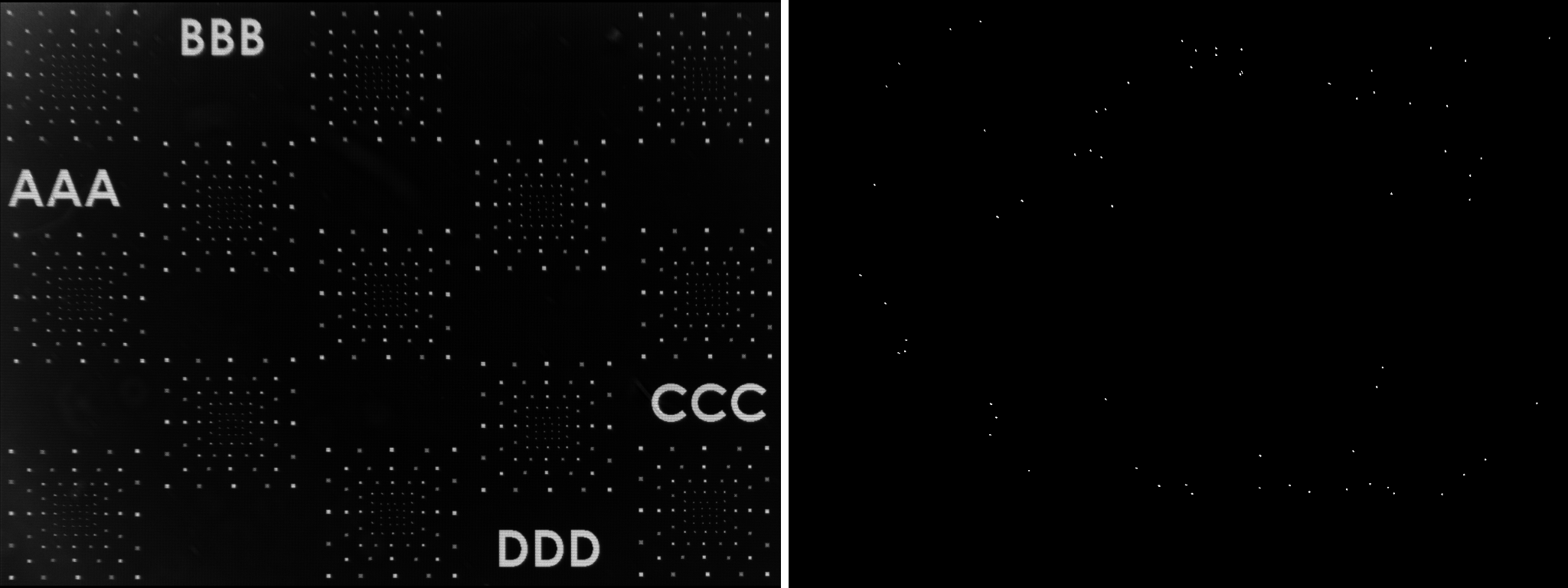} 
\setlength{\abovecaptionskip}{12pt}
\caption 
{ \label{fig:VibrationTestingDetection}
\textit{left} A reference focus pattern that is uploaded to the DMD. The 4$\times$4 set of mirrors arranged in a checkerboard pattern in the center of the square patterns are clearly visible. \textit{right} Malfunctioning micromirrors (on a device that was compromised during the re-windowing process) can be seen in a dark frame due to the light they scatter into the imaging system.} 
\end{figure}

The DMDs were inspected to detect any micromirrors that might have been tripped as a result of the mechanical tests. The DMDs were illuminated from the side at 24$^{\circ}$ with respect to the device normal and imaged at 0$^{\circ}$ (normal to the device) (Figure \ref{fig:VibrationTestingImagingSetup}). First, the inspection imaging system was focused on the DMD using a reference test pattern (Figure \ref{fig:VibrationTestingDetection}, \textit{left}). The test pattern consists of large and small features, which allow us to determine when the image is sufficiently focused. At the center of each square pattern is a set of 4$\times$4 mirrors flipped into opposite states in a checkerboard manner. When these single mirrors are visible, the system is sufficiently focused. Once focus is achieved, all of the mirrors are tilted away from the imaging system. When all mirrors are functioning, this configuration creates a black image. Any mirrors that are malfunctioning are clearly seen (Figure \ref{fig:VibrationTestingDetection}, \textit{right}). The number of failed micromirrors for each device (before and after the test) is shown in Table 6.

\begin{table}[ht]
\caption{Number of malfunctioned micromirrors for DMDs after the vibration testing.} 
\label{tab:FailedMIrrors}
\begin{center}       
\begin{tabular}{|c|l|c|c|} %% this creates two columns
%% |l|l| to left justify each column entry
%% |c|c| to center each column entry
%% use of \rule[]{}{} below opens up each row
\hline
\rule[-1ex]{0pt}{3.5ex}  Serial number & Window material & No. of tripped mirrors before  & No. of tripped mirrors after \\
\hline\hline
\rule[-1ex]{0pt}{3.5ex}  191207 & magnesium fluoride  & 2 & 2 \\
\hline
\rule[-1ex]{0pt}{3.5ex}  240207 & magnesium fluoride & 0 &  0\\
\hline
\rule[-1ex]{0pt}{3.5ex}  191306 & sapphire  & 0 & 0 \\
\hline
\rule[-1ex]{0pt}{3.5ex}  191005 & sapphire  & 0 & 0 \\
\hline
\rule[-1ex]{0pt}{3.5ex}  191105 & sapphire  & 0 & 0 \\
\hline
\rule[-1ex]{0pt}{3.5ex}  250305 & borosilicate  & 0 & 0 \\
\hline
\rule[-1ex]{0pt}{3.5ex}  250408 & borosilicate  & 0 & 0 \\
\hline
\end{tabular}
\end{center}
\end{table}
%----------------------------------------------------------------------------
% Vibration and Shock testing--->Package Integrity
%----------------------------------------------------------------------------
\subsection{Package Integrity}
After the vibration and mechanical shock tests, the integrity of the DMD packages was checked using the leak test procedure described in Section \ref{subsec:VibTest:LeakTesting}; the helium leak rates are given in Table \ref{tab:LeakTestAfter}.
\begin{table}[ht]
\caption{Leak rates for DMDs after the vibration testing.} 
\label{tab:LeakTestAfter}
\begin{center}       
\begin{tabular}{|c|l|c|l|} %% this creates two columns
%% |l|l| to left justify each column entry
%% |c|c| to center each column entry
%% use of \rule[]{}{} below opens up each row
\hline
\rule[-1ex]{0pt}{3.5ex}  Serial number & Window  material  & He leak rate [atm-cm$^3$/s] &  Notes \\
\hline\hline
\rule[-1ex]{0pt}{3.5ex}  190806 & fused silica &  6.2$\times$10$^{-7}$ &  stainless frame  \\
\hline
\rule[-1ex]{0pt}{3.5ex}  190805 & fused silica &  7.5$\times$10$^{-7}$ &  stainless frame    \\
\hline
\rule[-1ex]{0pt}{3.5ex}  191007 & fused silica &   7.0$\times$10$^{-7}$ &      \\
\hline
\rule[-1ex]{0pt}{3.5ex}  191006 & fused silica &   5.3$\times$10$^{-7}$ &      \\
\hline
\rule[-1ex]{0pt}{3.5ex}  191207 & magnesium fluoride &  9.3$\times$10$^{-8}$ &    \\
\hline
\rule[-1ex]{0pt}{3.5ex}  240207 & magnesium fluoride &  1.2$\times$10$^{-7}$ &    \\
\hline
\rule[-1ex]{0pt}{3.5ex}  191306 & sapphire &  2.2$\times$10$^{-7}$ &    \\
\hline
\rule[-1ex]{0pt}{3.5ex}  191005 & sapphire &   1.8$\times$10$^{-7}$ &   New AR coat \\
\hline
\rule[-1ex]{0pt}{3.5ex}  191105 & sapphire &   1.5$\times$10$^{-7}$ &   New AR coat \\
\hline
\rule[-1ex]{0pt}{3.5ex}  110306 & borosilicate &  7.5$\times$10$^{-10}$ & standard package   \\
\hline
\rule[-1ex]{0pt}{3.5ex}  250305 & borosilicate &  1.1$\times$10$^{-10}$ & standard package   \\
\hline
\rule[-1ex]{0pt}{3.5ex}  250408 & borosilicate &  7.1$\times$10$^{-10}$ & standard package   \\
\hline
\end{tabular}
\end{center}
\end{table}
\FloatBarrier
%%%%%%%%%%%%%%%%%%%%%%%%%%%%%%%%%%%%%%%%%%%%%%%%%%%%%%%%%%%%%%%%%%%%%%%%%%%%%
% Scattering measurements
%%%%%%%%%%%%%%%%%%%%%%%%%%%%%%%%%%%%%%%%%%%%%%%%%%%%%%%%%%%%%%%%%%%%%%%%%%%%%
\section{Reflectivity and Contrast Measurements}
\label{sec:ReflectivityAndContrast}
%----------------------------------------------------------------------------
% Reflectivity and Scattering Measurements--->Background
%----------------------------------------------------------------------------
\subsection{Background}
\label{subsec:Reflectivity:Background}
As described previously (Chapter \ref{subsec:Intro:MOS}), the individual spectral slits of a MOS can be defined using the individual micromirrors of a DMD. Each individual micromirror has four edges and a central via therefore there is the potential for incident light to be scattered in non-specular directions from micromirrors in both the \emph{off state} and \emph{on state}. Some part of this scattered light could reach the detector in the spectral arm and contribute to the background level, thereby increasing the background noise, thereby degrading the signal-to-noise ratio (SNR) of the spectra. For this reason the reflectance properties of the DMDs have been investigated and the amount of the scattered light assessed.\\

For 0.7\textsuperscript{$\prime \prime$} XGA devices the optical properties specified by TI\textsuperscript{TM} are:
\begin{itemize}
\item[-] Micromirror array fill factor: 92\%\cite{DMDdatasheet}. Fill factor is the fraction of a DMD pixel that is actually reflective. The specified micromirror pitch of 13.68 $\mu$m is the distance between the centers of the micromirrors, but there are vias in the centers of micromirrors and open gaps (about 0.6 $\mu$m wide) between adjacent micromirrors edges (Figure \ref{fig:DMD}) to allow rotational motion between the two operational stages without collision of the mirrors.  
\item[-] Micromirror array diffraction efficiency: 86\%\cite{DMDdatasheet}. This represents the loss from the specular reflection by diffraction coming from the finite size of the micromirrors.
\item[-] Micromirror surface reflectivity: 88\%\cite{DMDdatasheet}. The reflectivity of the micromirror surface determined by its material properties.
\item[-] Window transmission: 97\% for a single pass accounting for all losses at the front and back surface.\cite{DMDdatasheet}.
\end{itemize}
The DMD optical throughput parameters listed above were determined using a test configuration designed for simulating a projection system (i.e. a DLP\textsuperscript{\textregistered} system) over a wavelength range of 400-700 nm, an $f/3$ illumination beam, and an $f/2.3$ detector acceptance cone\cite{DMDdatasheet}. In a two arm MOS configuration the illumination will need to be slower than $f/2.35$ (corresponds to a 24$^{\circ}$ cone) to avoid optical overlap between the imaging and spectroscopy arms. For instance, the illumination in RITMOS\cite{Meyer2004} has the f-number of $7.62$, and in GMOX\cite{GMOXreport} and SAMOS\cite{SAMOS} - $f/4.1$. The slower optics has narrower acceptance cone for scattered light, which improves the contrast ratio \cite{DLPapplication}. In addition a broader spectral band, ranging from the UV to the IR is necessary for astronomical MOS applications. Therefore, the reflective and scattering properties of DMDs (with and without the protective cover windows) need to be measured using MOS relevant parameters as they could differ significantly from the values reported in the datasheet\cite{DMDdatasheet}. For this reason we performed measurements to assess the reflectance and the scattering properties of DMDs using both commercial spectrometers, and a custom test setup that simulates the optical configuration of a multi-object spectrograph.
%----------------------------------------------------------------------------
% Reflectivity and Contrast Measurements--->Reflectivity Measurement
%----------------------------------------------------------------------------
\subsection{Reflectance Measurements}
\label{subsec:Reflectance}
All measurements reflectance presented in this section were done on older DMDs - the SVGA devices with 868$\times$600 micromirrors with 17 $\mu$m micromirror pitch and 16 $\mu$m micromirror size. These devices have the fill-factor of 87\%\cite{FourspringPhD_Thesis} and a flip angle of 10$^{\circ}$.
The reflectivity of these DMDs was measured at NASA Goddard Space Flight Center (GSFC) using a Perkin Elmer (Model Lambda 950) spectrometer fitted with an Universal Reflectance Accessory (URA)\cite{URAperkin}. It is an accessory to measure specular reflectance of samples over a wide range of angles, which consists of a monochromator, a sample holder, and a detector on a rotational stage. Light of a chosen wavelength exits the monochromator and enters a chamber where the device under test is positioned on the sample holder at the center. The detector can be moved in a 360 degree range around the sample to measure the light intensity at any angle. The $f/\#$ of the illumination and the acceptance cone of the detector have the fastest setting of $f/10$ and $f/7.78$ respectively. This were the $f/\#$ settings used for the reflectance measurements described here.\\

The DMD was placed in the sample holder and angled at 8 degrees to the incoming light. The beam foot-print produced by the spectrometer is 3$\times$5 mm, which corresponded to 175$\times$294 micromirrors on the device. For these tests the DMD was not powered, and therefore all the mirrors were in the neutral (0$^{\circ}$) state, which have set the angle of specular reflectance at -8 degrees and this is where the detector was positioned.\\

\begin{figure}[tb]
\centering
\includegraphics[width=\textwidth]{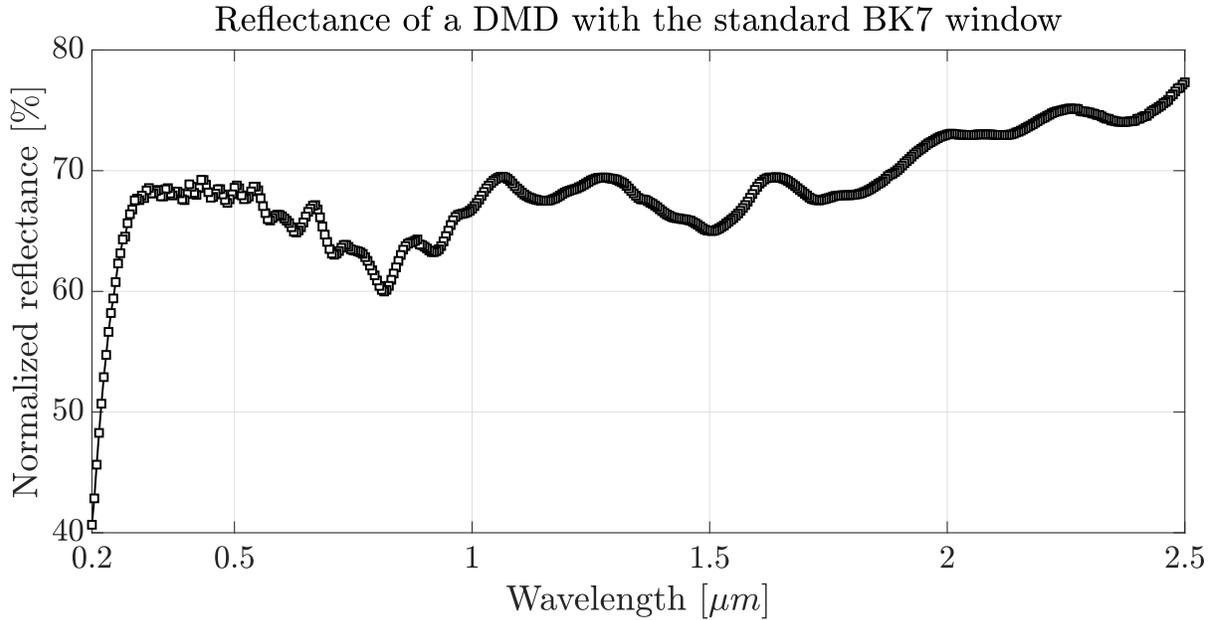} 
\setlength{\abovecaptionskip}{12pt}
\caption 
{ \label{fig:ReflectanceDMDborosilicate}
The reflectance of a DMD with the standard borosilicate protective window. Illuminated with an $f/10$ beam at an incidence angle of 8$^{\circ}$ and measured by a detector with a $f/7.78$ acceptance cone at the specular reflectance angle of -8$^{\circ}$.} 
\end{figure}

The raw reflected intensity (measured in arbitrary analog-to-digital units) was corrected for background signal, the source spectral intensity, and photodetector quantum efficiency (QE). The background signal was acquired by collecting the readings of the detector while the light source is switched off. The source spectral intensity was recorded by positioning the detector to be illuminated by the light coming from the monochromator directly. The calibration was done using the following formula:
\begin{equation}
\text{Reflectance}(\lambda) = \dfrac{\text{raw intensity}(\lambda) - \text{background}(\lambda)}{\text{source intensity}(\lambda) - \text{background}(\lambda)}
\end{equation}
where:
\begin{conditions}
 \lambda  & is the wavelength of light\\
 \text{source intensity} & direct wavelength-dependent signal from the monochromator on the detector\\
 \text{background} & signal on the detector when the light source is off.  
\end{conditions}
Using this approach we measured the reflectance of a DMD with the original borosilicate (Corning 7056\cite{DMDdatasheet}) protective window in place (Figure \ref{fig:ReflectanceDMDborosilicate}). The average reflectance we observed over the 0.2-2.5 $\mu$m range is presented in Table \ref{tab:Reflectance}. 

\begin{figure}[tb]
\centering
\includegraphics[width=\textwidth]{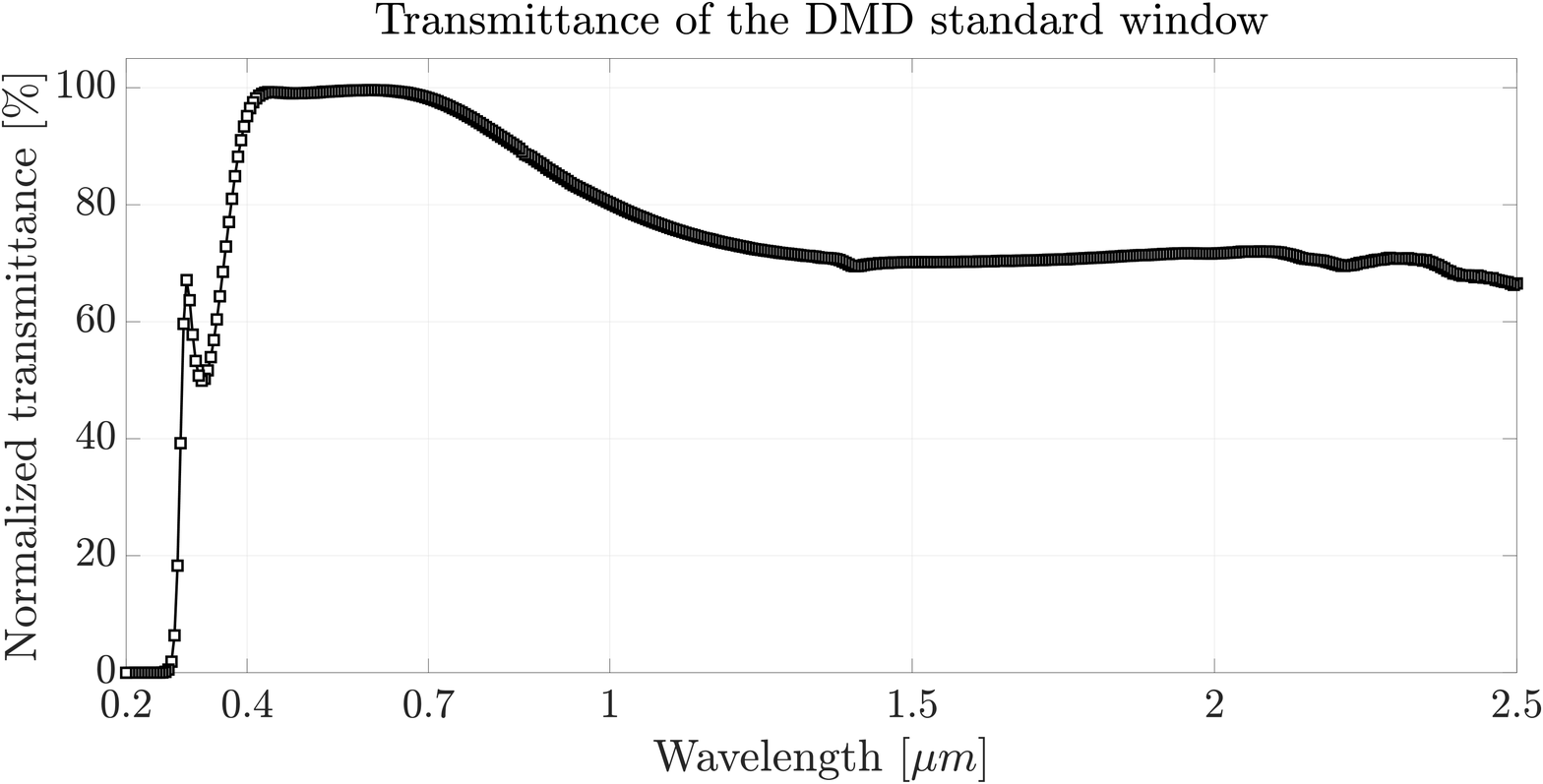} 
\setlength{\abovecaptionskip}{12pt}
\caption 
{ \label{fig:Transmittanceborosilicate}
Transmittance of the DMD standard window (Corning 7056) at normal incidence (measured after removed from the device). The window is optimized for the visible wavelengths (400-700 nm).} 
\end{figure}

\begin{figure}
\centering
\includegraphics[width=\textwidth]{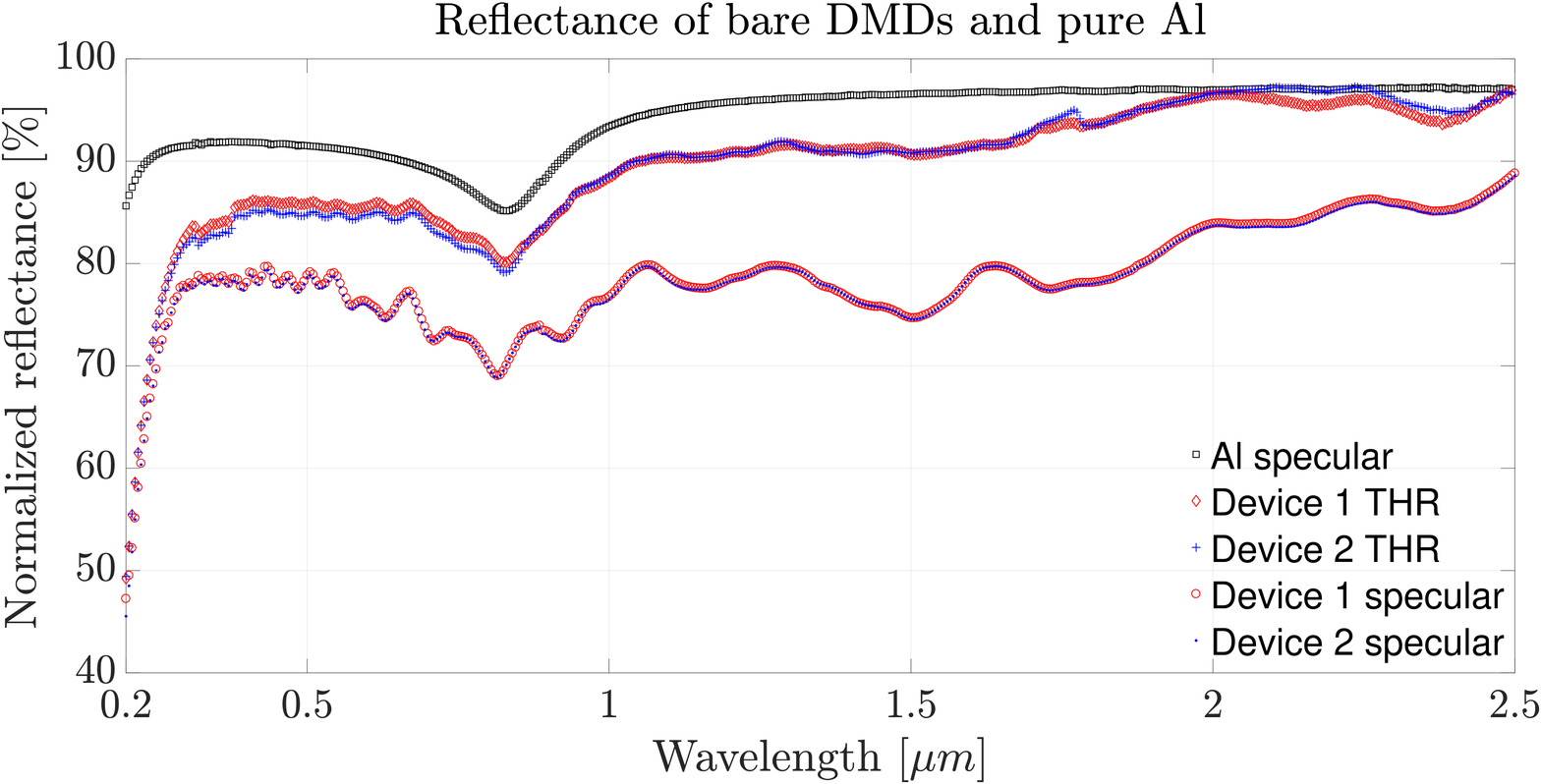} 
\setlength{\abovecaptionskip}{12pt}
\caption 
{ \label{fig:ReflectanceBareDMDsVSaluminum}
Specular reflectance and total hemispherical reflectance (THR) of two bare DMDs, corrected for the 87\% fill-factor \cite{FourspringPhD_Thesis} to compare with the reflectance of pure Al (also shown).} 
\end{figure}

\begin{figure}[tb]
\centering
\includegraphics[width=\textwidth]{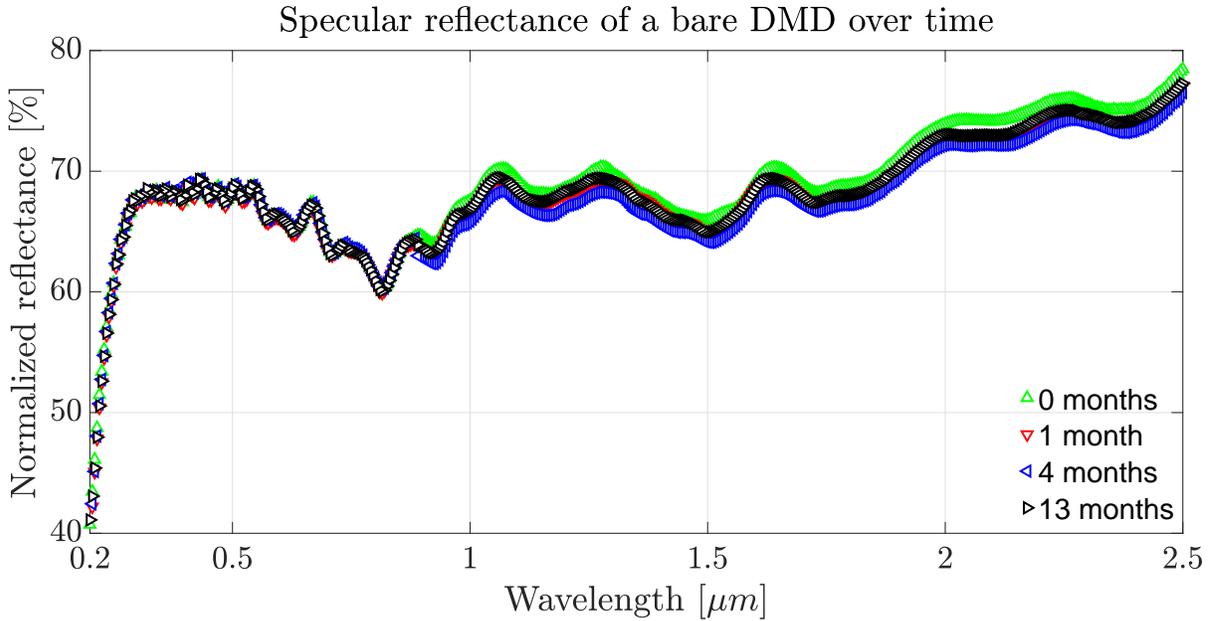} 
\setlength{\abovecaptionskip}{12pt}
\caption 
{ \label{fig:ReflectanceBareDMD}
Specular reflectance of one DMD after the protective window has been removed as a function of wavelength, measured with Perkin Elmer Lambda 950 Spectrometer with a URA\cite{URAperkin} with 8$^{\circ}$ angle of incidence. The “0 month” trace corresponds to data taken immediately after window removal. The same measurement was repeated after 1, 4, and 13 months.} 
\end{figure}

\begin{table}[htb]
\caption{Average reflectance of a DMD (presented on Figure \ref{fig:ReflectanceDMDborosilicate}) by wavelength range.} 
\label{tab:Reflectance}
\begin{center}       
\begin{tabular}{|l|c|c|c|c|}
\hline
\rule[-1ex]{0pt}{3.5ex}  Wavelength range [$\mu$m] & 0.2 - 2.5 & 0.2 - 0.4  & 0.4 - 0.75 & 0.75 - 2.5 \\
\hline
\rule[-1ex]{0pt}{3.5ex}  Reflectance & 68.4\%  & 63.1\% & 66.6\% & 69.3\% \\
\hline
\end{tabular}
\end{center}
\end{table}
This average reflectance represents the absolute throughput of the device and it includes all the losses from diffraction, scattering, and the fill-factor of the device described the Section \ref{subsec:Reflectivity:Background}. These data show a reflectance drop below 325 nm that is due to transparency loss of the standard borosilicate window (Figure \ref{fig:Transmittanceborosilicate}) used by TI\textsuperscript{TM} in packaging of the DMDs.
Therefore, using off-the-shelf DMDs as light modulators for applications requiring wavelengths below 325 nm is extremely ineffective. The only way to avoid this issue is by replacing the TI\textsuperscript{TM}-provided windows with a UV-transmissive substrate (e.g. Magnesium Fluoride or UV-grade sapphire\cite{Travinsky_Rad_test_Journal_I}).\\
\\
In order to be able to choose optimal replacement windows the intrinsic reflectance properties of the DMD micromirrors themselves must be measured. To do this the protective window was removed and the reflectance of the bare device over the 200-2000 nm range measured using the same Perkin Elmer spectrometer fitted with the URA\cite{URAperkin} (Figures \ref{fig:ReflectanceBareDMDsVSaluminum} and \ref{fig:ReflectanceBareDMD}).
The reflectance values were compensated for the losses associated with the non-reflective parts of the micromirror array - the gaps between the micromirrors and the via (Figure \ref{fig:DMD}) by scaling the absolute reflectance by the 87\% fill-factor\cite{FourspringPhD_Thesis}. We observed that  the scaled absolute reflectance values of the bare DMDs are still lower than expected for aluminum-based mirrors ($>$85\% over the wavelength range of 400-700 nm \cite{EdmundAluminum}). We compared the fill-factor corrected specular reflectance of two bare DMDs to the specular reflectance of pure aluminum (Figure \ref{fig:ReflectanceBareDMDsVSaluminum}). 
\\
The corrected reflectance of the two DMDs (curves labeled as ``specular'') are  lower than that of the aluminum sample, especially for shorter (200-300 nm) wavelengths. Part of these reflectance losses could be also attributed to diffraction, as discussed in Chapter \ref{subsec:Reflectivity:Background}. In order to account for these losses, we also performed Total Hemispherical Reflectance (THR) measurement, by using a 60 mm integrating sphere accessory attached to the Perkin Elmer spectrometer. We used a NIST traceable standard\cite{NISTstandard} to calibrate out any losses related to the integrating sphere itself. The results of the traces in Figure \ref{fig:ReflectanceBareDMDsVSaluminum} indicate that the THR values are indeed higher than the ones corresponding to the specular component alone. However, even after accounting for the diffraction losses, the THR data still do not match the results obtained from a pure aluminum sample. The overall reduced reflectance of the these devices (compared to pure aluminum) could be reasonably attributed to the fact that the aluminum on the DMD micromirorrs may not be pure, but rather based on some type of alloy, and could possibly contain small amount of titanium (Ti) and copper (Cu). This addition to pure aluminum could be used to increase the mechanical strength of the micromirrors while minimizing the loss of aluminum’s intrinsic high reflectivity\cite{LanseThesis}. The fact that the DMD mirrors are coated with some type of proprietary alloy with high aluminum content was also confirmed by TI\textsuperscript{TM}\cite{RickOden}. \\

While replacing the original window of a DMD with another substrate the micromirror array could be exposed to ambient conditions. This can potentially oxidize the micromiror surface and change its reflectance properties. To investigate this potential change in the reflectance properties of DMDs we removed the protective window from one DMD and monitored the reflectance of this bare device over the period of 13 months, using the same setup with the Perkin Elmer Lambda 950 spectrometer with URA.(Figure \ref{fig:ReflectanceBareDMD}). We did not detect measurable changes in the reflectance properties of the DMD over the period of 13 months.

\FloatBarrier
%----------------------------------------------------------------------------
% Reflectivity and Scattering Measurements--->Reflectivity Measurement--> coating of DMD active area
%----------------------------------------------------------------------------
\subsection{Coating of DMD active area}
The low reflectance of DMD micromirrors below 0.3 $\mu$m (even for bare devices) limits the use of DMDs as programmable slit masks in a spectrograph designed for use in the deep UV. One means to improve the DMD deep UV reflectivity is to coat the active area of the DMD with fresh aluminum. In order to test this concept we carried out a coating task of depositing an aluminum layer with the thickness between 50 - 60 nm on the Device \#1 whose initial (i.e. before this fresh coating) reflectance is shown in Figure \ref{fig:ReflectanceBareDMDsVSaluminum}.\\

A mask was manufactured with the proper dimensions so that it was placed on the device during the coating process to protect the wire bonds from being coated and resulting in an electrical short. This mask has an aperture that was located at the center of the DMD and matched in size the form factor (868$\times$600 pixels) of the DMD active area. The DMD with covering mask was then placed inside a high-vacuum coating chamber where the aluminum (Al) deposition would take place.\\

The Al coating process used the Physical Vapor Deposition (PVD) method where the Al is placed in a resistive bowl (made out of Tungsten) with electrical wires attached to it. A current is passed through this bowl until the Al melts and evaporates onto the DMD, which is placed directly on top of the bowl inside the vacuum chamber.\\ 

One of the parameters that determine the quality of the deposited Al is that a high vacuum is maintained during the coating process. In our case this was between 10$^{-8}$ Torr to 10$^{-7}$ Torr. Another parameter that was controlled was to ensure the absence of residual water vapor and oxygen during growth. The presence of these substances can reduce the quality of the deposited aluminum through oxidization. Finally the Al deposition rate can also affect the layer quality and it was maintained at 100 $\AA$ per second or higher, in order to have a dense aluminum film.\\

\begin{figure}
\centering
\includegraphics[width=\textwidth]{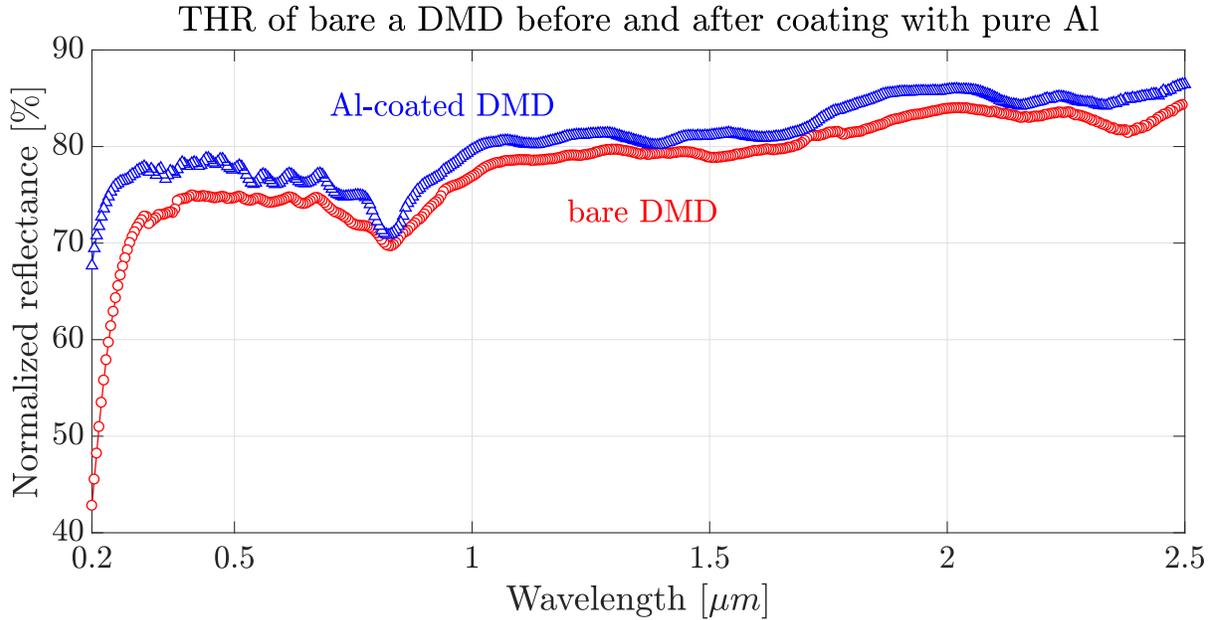} 
\setlength{\abovecaptionskip}{12pt}
\caption 
{ \label{fig:ReflectanceBareDMDVSaluminum2}
Total hemispherical reflectance (THR) of a DMD before and after coating with pure aluminum (layer thickness 50-60 nm). The coating resulted in an increase in reflectance of DMD micromirrors from 42\% to 68\% at 0.2 $\mu$m.} 
\end{figure}

The coated DMD (Figure \ref{fig:ReflectanceBareDMDVSaluminum2}) shows a substantial improvement in reflectance reaching a value of 68\% at 0.2 $\mu$m, when compared to the value prior to the coating application (42\%). The results also show a more modest (4-5\%) increase in the 0.4-2.5 $\mu$m wavelength range (Table \ref{tab:ReflectanceCoated}). This result agrees with the results described in the Section \ref{subsec:Reflectance} that the Al from which the DMD micromirrors are made from is in fact some type of alloy that exhibits poorer reflectance properties than the pure aluminum.\\
\begin{table}[htb]
\caption{Average reflectance values of a bare DMD (presented on Figure \ref{fig:ReflectanceBareDMDVSaluminum2}) before and after coating with aluminum.} 
\label{tab:ReflectanceCoated}
\begin{center}       
\begin{tabular}{|l|c|c|c|c|}
\hline
\rule[-1ex]{0pt}{3.5ex}  Wavelength range [$\mu$m] & 0.2 - 2.5 & 0.2 - 0.4  & 0.4 - 0.75 & 0.75 - 2.5 \\
\hline
\rule[-1ex]{0pt}{3.5ex}  THR reflectance of bare DMD& 77.9\%  & 67.3\% & 74.3\% & 79.9\%\\
\hline
\rule[-1ex]{0pt}{3.5ex}  THR reflectance of coated DMD& 80.65\%  & 76.1\% & 76.9\% & 81.9\%\\
\hline
\end{tabular}
\end{center}
\end{table}

A topic of future investigation will be to determine whether any of the evaporated aluminum becomes deposited on the substrate behind the DMD mirrors by passing through the gaps between mirrors. This may cause electrical problems that could affect the electromechanical mechanism that tilts the micromirrors. Additionally, the lifetime of a coated DMD might be compromised because the passive environment, which is present in standard TI DMD hermetically sealed packages, no longer exists. However, there are TI DMDs that are sold not hermetically sealed and still operate for extended periods of time, for instance, in TI spectrometers. Additionally, we have been successful in re-windowing hermetically sealed TI DMDs with other types of glasses\cite{Travinsky_Rad_test_Journal_I}. The re-windowed devices remain operational today, some about two years after re-windowing, although there was no effort to restore the original environment in re-windowed packages. Future investigative effort will be put into further assessment of the degree of viability of this approach of re-coating DMD micromirrors with evaporated aluminum, and the results obtained so far are very promising.\\
\FloatBarrier
%-------------------------------------------------------------------------
% Reflectivity and Contrast Measurements--->Contrast Measurement
%-----------------------------------------------------------------------
\subsection{Contrast Measurements}
\label{subsec:Contrast}
In a MOS system a measure of performance is the ability to both direct as much light as possible to the spectrometer when a particular target is selected and to not direct any of that target's light to the spectrometer when it is not selected. This ability depends not only on the intrinsic reflective/scattering properties of the slit mask, but also on the $f/\#$ of the illuminating and collecting optics. For instance, the optical parameters presented in the TI data sheet\cite{DMDdatasheet} for this DMD type were acquired by illuminating a DMD with $f/3$ and collecting with $f/2.4$ (p. 20 in the data sheet). This is typical for projection systems, but not for MOS. It is important therefore to measure the contrast ratio of DMDs under MOS-specific conditions. In this paper the results of two measurement approaches to assess the contrast ratio of the 0.7\textsuperscript{$\prime \prime$} XGA DMDs in a typical MOS configuration are presented.\\ 

\begin{figure}[htb]
\centering
\includegraphics[width=\textwidth]{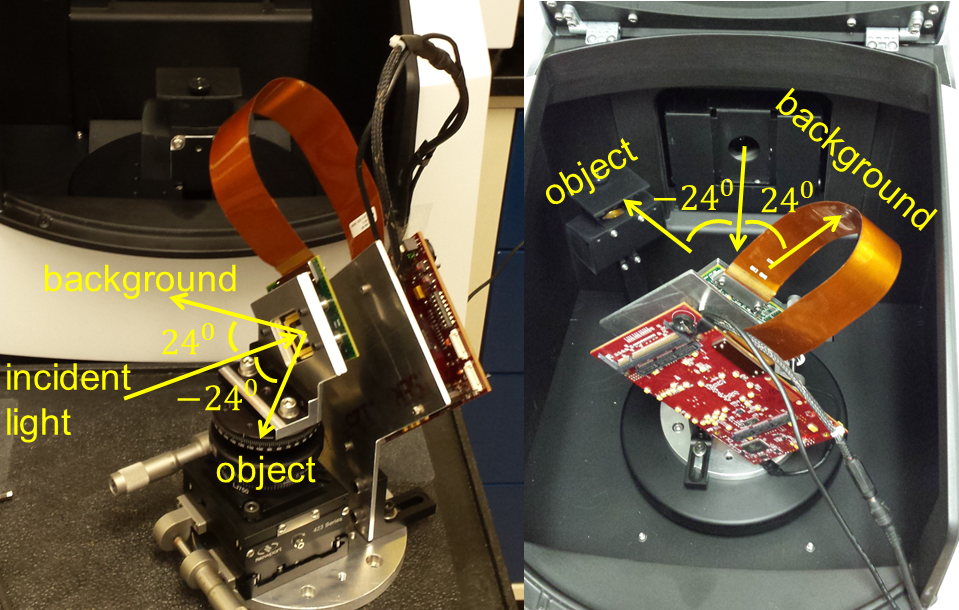} 
\setlength{\abovecaptionskip}{12pt}
\caption 
{ \label{fig:Cary5000setup}
\textit{left hand side}: DMD set up for contrast measurement outside the Universal Measurement Accessory (UMA) of Cary 5000 spectrophotometer. The directions of illumination and specular and anti-specular reflections are marked as object and background respectively. \textit{right hand side} The DMD is illuminated from the normal to the surface of the device direction, with all mirrors tilted into the \emph{on state} (-12$^{\circ}$) position. The specular reflection was measured at the location of the object (the specular direction, -24$^{\circ}$) and the scattered light was measured at the location of the background (the anti-specular direction, 24$^{\circ}$).} 
\end{figure}

\begin{figure}[tb]
\centering
\includegraphics[width=\textwidth]{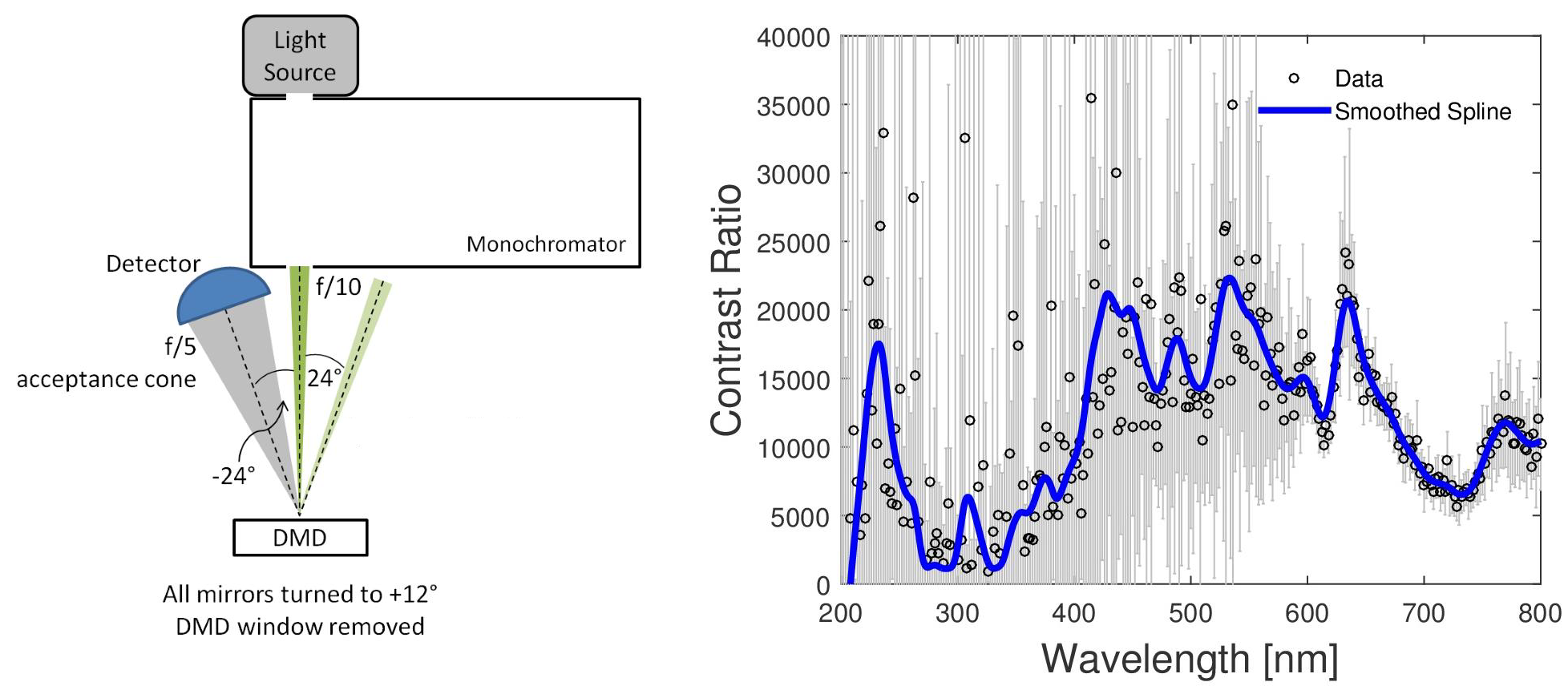} 
\setlength{\abovecaptionskip}{12pt}
\caption 
{\label{fig:Cary5000results}
\textit{Left hand side}: The contrast measurement performed using the Cary 5000 spectrophotometer with the Universal Measurement Accessory. All mirrors were tilted into the \emph{on state} (-12$^{\circ}$) position and the signal was measured at the specular and anti-specular directions, at -24$^{\circ}$ and 24$^{\circ}$, respectively. \textit{Right hand side}: The contrast is the ratio of the signal measured at -24$^{\circ}$ vs. 24$^{\circ}$ (the \emph{on state} and the \emph{off state}, respectively). The large uncertainty in the measurement is due to poor sensitivity of the detector at these signal levels.} 
\end{figure}

The first method utilized a Agilent Cary 5000\cite{Cary5000} spectrophotometer with a Universal Measurement Accessory (UMA)\cite{Cary5000UMA}, which consists of a monochromator, a sample holder, and a detector that can be positioned at any angle around the sample holder. Light of a chosen wavelength from the monochromator illuminates the DMD installed in the sample holder with an $f/10$ cone and the reflected light is collected by the detector with an $f/5$ acceptance cone (Figure \ref{fig:Cary5000setup}). To permit the measurement of the contrast ratio in the wavelengths outside the useful range of the standard TI\textsuperscript{TM}-provided protective window, the window was removed.  The DMD mirrors were tilted to the -12$^{\circ}$ direction. The plotted contrast ratio is the ratio of the signal measured by the detector at the specular direction (-24$^{\circ}$) to that measured in the anti-specular direction (24$^{\circ}$) (Figure \ref{fig:Cary5000results}. The measured contrast ratios display large uncertainty bars because of the poor sensitivity of the photodiode, even when using the maximum integration time, for the measurement of the scattered light in the anti-specular direction. The contrast measured in this configuration was at least 5000:1 for wavelengths longer than 0.4 $\mu$m.\\
\\
\\

\begin{figure}[tb]
\centering
\includegraphics[width=\textwidth]{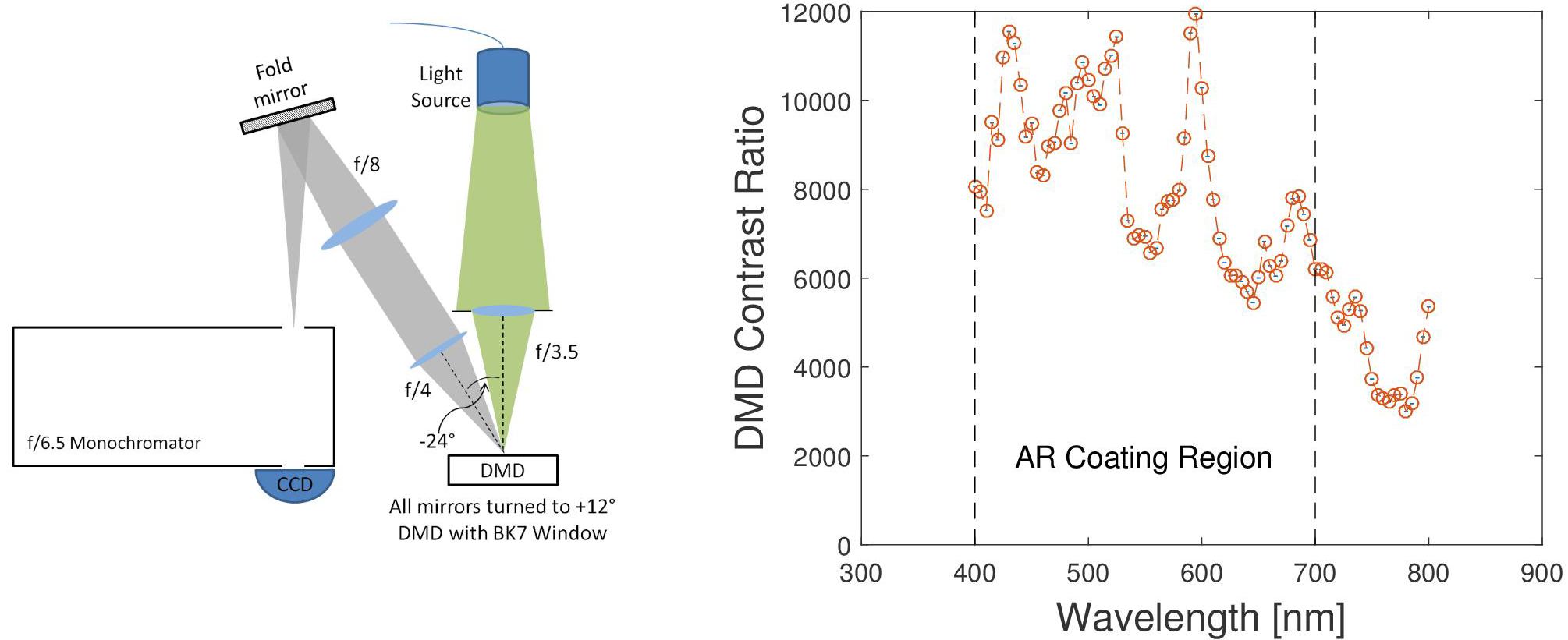} 
\caption 
{ \label{fig:RITsetupResults}
\textit{Left hand side}: The contrast measurement performed in a way that simulates the configuration of an actual DMD-based MOS. The light from a broadband source is focused onto the DMD under test with an $f/3.5$ lens, resulting in a 300 $\mu$m diameter spot ($\sim$22 micromirrors). This spot is re-imaged with 2:1 magnification onto the entrance slit of the monochromator. The monochromator disperses the light coming from the DMD into its spectral content. A detector placed at the output port of the monochromator registers the wavelength-dependent signal.  \textit{Right hand side}: The DMD contrast ratio is the ratio of the signal on the detector when the DMD mirrors are in the \emph{on state} (-12$^{\circ}$) to the signal on the detector when all the DMD mirrors are in the \emph{off state} (+12$^{\circ}$). The dotted lines signify the optimal wavelength range of the protective window on the DMD under test.} 
\end{figure}

The second method of measuring the DMD contrast involved assembling an optical system more similar to that found in a MOS configuration (Figure \ref{fig:RITsetupResults}, \textit{left hand side}). This measurement was performed on a device with the standard protective window on. Light from a broadband source was focused by a 180 mm $f/3.5$ lens onto the DMD under test, creating a 300 $\mu$m diameter spot, corresponding to 22 DMD mirrors. The reflected and the scattered light was collected and collimated using an $f/4$ lens at -24$^{\circ}$, by turning the DMD mirrors into the \emph{on state} (-12$^{\circ}$) and the \emph{off state} (12$^{\circ}$), respectively. The collimated light was focused by a second $f/8$ lens and sent into an $f/6.5$ monochromator to scan through the wavelengths. An imaging sensor was used at the output port of the monochromator and the exposure times were adjusted to maintain a high signal-to-noise ratio.\\

The contrast ratio measured with this second setup was greater than 5000:1 in the spectral region where the protective borosilicate window is transparent. This agrees with the first set of measurements, obtained with the Cary 5000 spectrophotometer. A system designed with stray-light reduction in mind (e.g. usage of optical baffles and absorptive coatings) can further reduce the amount of scattered light in the spectral channel.\\
\\
\FloatBarrier
%%%%%%%%%%%%%%%%%%%%%%%%%%%%%%%%%%%%%%%%%%%%%%%%%%%%%%%%%%%%%%%%%%%%%%%%%%%
% Low Temperature Testing
%%%%%%%%%%%%%%%%%%%%%%%%%%%%%%%%%%%%%%%%%%%%%%%%%%%%%%%%%%%%%%%%%%%%%%%%%%%
\section{Low Temperature Testing}
\label{sec:LowTempTest}
In this section we present the results of tests to assess DMD operability at cryogenic temperature.
%-------------------------------------------------------------------
\subsection{Introduction}
\label{subsec:LowTempTest:Intro}
For applications where DMDs are used as programmable slit-masks in spectrographs operating in the near-infrared regime, thermal emission is a significant design consideration.  Emission is typically mitigated by cooling hardware seen by the detector to cryogenic temperatures, which greatly reduces self-emission.  Hence, for near-infrared applications the operability of DMDs at cryogenic temperature is of great interest. Here we present preliminary results from tests conducted at Johns Hopkins University in Baltimore, MD, on the operability and lifetime of DMDs at cryogenic temperature.

\subsection{The Experiment}
\label{subsec:LowTempTest:Experiment}
A Texas Instruments 0.7\textsuperscript{$\prime \prime$} XGA DMD was cooled in a thermal vacuum chamber and optically tested to assess the operability of the device on a mirror-by-mirror basis.  The device was controlled by a formatter board supplied with the DLi 4120 development kit procured from Digital Light Innovations.\\

To evaluate DMD operability, a direct imaging technique was established with the DMD being positioned inside a vacuum chamber facing a viewport window, while a high resolution digital camera (Basler acA2040-90) outside the chamber, and at normal incidence to the array, produced video framerate images of the mirror surface.  An illumination source, also outside the chamber, consisting of an array of LEDs behind a diffuser, was oriented 24$^{\circ}$ off normal incidence to the DMD so as to reflect light from those mirrors tilted 12$^{\circ}$ toward the source, while mirrors tilted away from the source reflected no light into the camera. With this setup, the resolution of the camera was sufficient to discern the change in state of a single mirror.\\

\begin{figure}[tb]
\centering
\includegraphics[width=\textwidth]{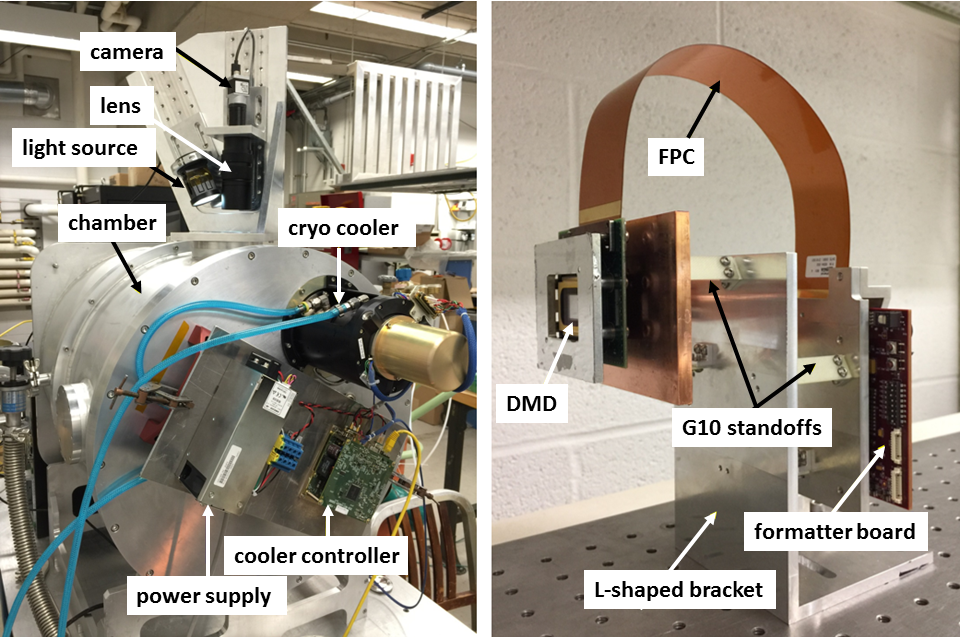} 
\setlength{\abovecaptionskip}{12pt}
\caption 
{\label{fig:DMD_CryoTest}
\textit{Left hand side}: Photograph showing the test chamber and optical setup used for cryogenic DMD operability testing. \textit{Right hand side}: Photograph showing the DMD and formatter board mounted to the chamber interface bracket.} 
\end{figure}

The DMD was cooled using a Sunpower Cryotel GT closed-cycle cryogenic cooler mounted to the lid of the chamber in close proximity to the DMD.  See Figure~\ref{fig:DMD_CryoTest} for photographs of the chamber test setup, as well as the DMD and formatter board mounting configuration. Two copper thermal straps connecting the cooler tip to the DMD thermal interface, i.e. a copper base plate, conducted away the radiative heat load on the DMD, as well as a significant parasitic heat load (of order 8 Watt) from the FPC (flexible printed circuit), which connects the DMD board to the ambient temperature formatter board that drives it. Here the cryogenic cooler serves to only cool the DMD, hence the rest of the chamber internals, including the formatter board, are kept at ambient temperature. To achieve this thermal condition, the DMD and its copper mounting plate were isolated from its chamber interface, a right-angle bracket, by three G10 standoffs. The formatter board was mounted on the opposite side of the right-angle bracket, and was also isolated by G10 standoffs, with a thermal strap connecting it to the chamber lid to passively remove heat generated by the board during operation.\\  

Temperatures of the DMD and formatter board were monitored using temperature diodes (Lakeshore model DT-670), with the cooler power being adjusted to achieve the desired DMD temperature. Given the large parasitic heat load through the FPC, the lowest achievable temperature for this test configuration was 78~K, sufficiently close to our goal of 77~K.\\

To assess operability of each mirror in the array, image processing analysis was performed on image frames from the digital video camera with the DMD array set to either the \emph{off state} (i.e. black image) or the \emph{on state} state (i.e. white image).  A single mirror stuck in the on state shows white against a black background when the array is set to the off state. Similarly, a single mirror stuck in the off state shows black against a white background when the array is set to the on state.  

To test this approach, and to validate the Matlab image processing algorithm developed for this study, state patterns consisting of intentionally generated stuck mirrors, either white or black, were uploaded to the DMD with and otherwise black or white (respectively) background. The software successfully detected the simulated stuck mirrors at an accuracy of 99.62\% over 100 trials of black and white frames. There were zero false positives.

\subsection{Results}
\label{subsec:LowTempTest:Results}
Basic operatiblity tests were conducted on the DLP7000 at 78~K.  Image analysis results showed zero stuck mirrors.  However, one frame was captured that showed an entire row of micromirrors in the \emph{on state} state against the \emph{off state} background.  This anomaly occurred only once in all of the frames captured.\\
Additionally, an accelerated lifetime test was conducted to evaluate the operability of the DLP7000 at cryogenic temperature.  The DMD mirror array was cycled between the \emph{on state} and the \emph{off state} at a rate of 100~Hz for a total of 200,000 flips; 200,000 flips being a very rough estimate as to the number of state changes expected over the 10 year life of a multi-object spectrograph utilizing DMD technology. Image processing analysis before and after the lifecycle testing showed zero mirror failures.  It should be noted that the parasitic heat load from the formatter board increased due to self heating during the lifetime test, driving the DMD temperature to 90~K over at the completion of the test, which had a duration of approximately 30 minutes.

% Discussion and Conclusions
%%%%%%%%%%%%%%%%%%%%%%%%%%%%%%%%%%%%%%%%%%%%%%%%%%%%%%%%%%%%%%%%%%%%%%%%%%%%%
\section{Discussion and Conclusions}
\label{sec:Conclusions}
%----------------------------------
 \subsection{Radiation testing}
%---------------------------------------
We performed two separate experiments to test DMD reliability under heavy ion irradiation. There were no hard failures or any other permanent damage, even under unrealistically high heavy ion fluxes. All upset mirrors were instantly cleared by uploading a new pattern onto the DMD. Our results allow to conclude that commercially available digital micromirror devices have extremely limited sensitivity to non-destructive heavy-ion induced state changes. All single event upsets can be cleared by a remote soft reset. We calculated an expected in-orbit micromirror upset rate to be about 5.6 micromirrors/day for the worst week scenario in interplanetary space under standard space craft shielding. This suggests that under those conditions a DMD-based instrument would experience only negligible heavy-ion induced single event upset rate burden. 
%----------------------------------
 \subsection{Vibration and shock testing}
%---------------------------------------
To investigate the reliability of individual micromirrors in the 0.7\textsuperscript{$\prime\prime$} XGA DMDs, we performed mechanical tests along X, Y, and Z directions and imaged the DMDs to search for failed mirrors. The DMDs were accelerated with a frequency spectrum from 20 Hz to 2 kHz, with an rms acceleration of  14 g and subjected to a shock response spectrum of 100 Hz - 10 kHz, with accelerations 500 g. We found no mirrors that have been tripped or failed as a result of the mechanical tests.\\

We also investigated the hermeticity of the standard TI\textsuperscript{TM} 0.7\textsuperscript{$\prime\prime$} XGA DMD package, which is hermetically sealed, before and after the vibration and mechanical shock tests. We measured helium leak rates on the order of few 10$^{-10}$ atm-cm$^3$/s for these standard-packaged devices. We also evaluated the hermeticity of re-windowed devices with epoxied packages. These exhibited helium leak rates of few 10$^{-7}$ atm-cm$^3$/s, close to the MIL-STD-883 limit for hermeticity (for these devices) of 1.4$\times$10$^{-10}$ atm-cm$^3$/s. The leak rates did not change significantly after the vibration and shock tests for any of the package types.\\

%In summary, the vibration and shock tests described in this work did not adversely affect the performance of DMDs or the custom packages with UV and IR transmissive windows. These GEVS tests suggest that DMDs are extremely robust and not sensitive to the potential vibroacoustic environments experienced during launch.
%----------------------------------
 \subsection{Reflectance and Contrast Ratio Measurements}
%---------------------------------------
We performed measurements of specular and total hemispherical reflectance (THR) on TI\textsuperscript{TM} 0.7\textsuperscript{$\prime\prime$} XGA DMDs with their protective windows in place and removed to estimate the throughput, durability, and scattering properties of the micromirrors reflective area. We found that the reflectivity of bare devices is very stable and does not change after being exposed to ambient atmospheric conditions over a period of 13 months. However, the base reflectivity of these devices is relatively low ($\sim$68\% over the 0.2 - 2.5 $\mu$m range).\\

We also measured the contrast ratio of 0.7\textsuperscript{$\prime \prime$} XGA DMDs in two MOS-like configurations. We found that contrast ratios of 6000:1 - 10000:1 can be reached in an $f/4$ spectrograph, depending on the wavelength of light and the quality of the DMD window and anti-reflection coatings.\\

\subsection{Coating of DMD active area}
We performed a coating experiment, where the active area of one DMD was coated with a fresh layer of pure aluminum (thickness $\sim$ 60 nm). We observed the average normalized reflectance of the newly coated device increased from an average of $\sim$ 68\% to an average of 80\% in the 0.2-2.5 $\mu$m spectral region. This boost in reflectance along with the replacement of the standard borosilicate window with one of the three substrate options mentioned above, will provide a viable path to use the DMD as a programmable slit mask in the UV or IR spectral range. Future tasks will include verifying that aluminum-coated DMD will remain fully functional.\\

%----------------------------------
 \subsection{Low Temperature Testing}
%---------------------------------------
The operability of a single 0.7\textsuperscript{$\prime \prime$} XGA DMD at low temperatures, down to very near the temperature of liquid nitrogen, was evaluated.  Both basic operability and lifetime tests were conducted.  Aside from one minor anomaly, the device worked flawlessly during basic operations,  and upon completion of a lifetime test consisting of 200,000 mirror flips, there were zero defective mirrors.  Overall, the results indicate that the devices are insensitive to these low temperatures, which bodes well for both ground-based and space-based applications requiring cryogenic operation. One issue uncovered during these tests that is of significance is the high parasitic heat load between the formatter board and DMD. For cryogenic applications, such as IR spectrography, the roughly 8 Watt load is substantial. This could be largely eliminated if the cable length were increased substantially, which is possible only with custom electronics, as the impedance of the FPC supplied with the DLi 4100 development kit is matched to the drive frequency of the stock board. An alternative approach would be to develop electronics that operate at cryogenic temperature.\\
Future experiments will concentrate on investigating the behavior of TI-packaged and re-windowed DMDs in cryogenic temperatures for extended periods of time with extremely low micromirror flip duty cycle (e.g. flip the micromirrors once every half an hour, while keeping the device cooled for several days and longer), as this would be more ``MOS-like'' DMD operation.

%----------------------------------
 \subsection{Summary}
%---------------------------------------
We performed a series of tests to assess the suitability of Digital Micromirror Devices (DMDs) for usage as programmable slits in ground-based and space-based Multi Object Spectrometers (MOS). We found that DMDs are extremely reliable, are not affected by vibrations and mechanical shock at launch, are not sensitive to heavy-ion radiation, insensitive to cryogenic operation, and possess good reflectance properties with potential for enhancement. We also found that current generation of DMDs have low scattered light levels, that can be even further reduced by application-driven optical system design. Our findings confirm that DMDs are well positioned for use in astronomical multi-object spectrographs on ground-based and space-based telescopes. 

\appendix 
% \disclosures 
\subsection*{Disclosures}
The authors have no relevant financial interests in the manuscript, and no other potential conflicts of interest to disclose.

\acknowledgments 
The authors acknowledge that this research was supported by a NASA Strategic Astrophysics Technology grant number NNX14AI62G. We thank Vladimir Horvat and Bruce Hyman from Texas A\&M University Cyclotron Institute for providing their prompt assistance during the testing and answering practical questions about the TAMU Cyclotron facility. Part of this work was performed at NASA Goddard Space Flight Center; we would like to thank the team at Sierra Lobo Vibration Facility for their help and guidance during the vibration testing. Also Linette Kolos and Felix Threat [in NASA-GSFC/552] for their support in performing aluminum deposition on the bare DMD. We also thank Michael Douglass, Benjamin Lee, and Patrick Oden from Texas Instruments Incorporated for productive discussions about the performance of DMDs under extreme conditions. In addition, we are thankful to Mike Buffalin and John ``Sean" Greenslade from ``The Construct at RIT'' for sharing their expertise in rapid manufacturing, and help in designing custom parts for the test setups. ZN acknowledges the hospitality of the University of Toronto Department of Astronomy and Astrophysics during the time this paper was completed.

%%%%% References %%%%%

\bibliography{references}   % bibliography data in references.bib
\bibliographystyle{spiejour}   % makes bibtex use spiejour.bst

%%%%% Biographies of authors %%%%%

\vspace{2ex}\noindent\textbf{Anton Travinsky} is a doctoral student at Rochester Institute of Technology. He received his B.Sc. degree in Mechanical Engineering from the Technion - Israel Institute of Technology (Haifa, Israel) in 2010 and M.Sc. in Electrical Engineering from the RWTH Aachen University (Aachen, Germany) in 2014. His current research interests include MOEMS-based optical and optomechanical systems, and reliability of MOEMS in extreme conditions. He is a student member of SPIE.\\
\vspace{2ex}\noindent\textbf{Zoran Ninkov} is a professor in the Center for Imaging Science (CIS) at the Rochester Institute of Technology. He completed his BSc in physics at the University of Western Australia, MS in chemistry at Monash University, and his PhD in astronomy at the University of British Columbia. He was a postdoctoral fellow at the University of Rochester working on infrared detector array for the Spitzer Space Telescope. His research interests are in detectors and instrumentation.\\
\vspace{2ex}\noindent\textbf{Dmitry Vorobiev} is a post-doctoral researcher in the Center for Imaging Science (CIS) at the Rochester Institute of Technology. He completed his BSc in Astronomy at the University of New Mexico, as well as an MS in Imaging Science and PhD in Astrophysical Sciences and Technology at RIT. Dmitry is interested in the development and characterization of instrumentation for astronomy and remote sensing, particularly devices that measure the polarization and spectrum of light.
\vspace{1ex}
\noindent Biographies and photographs of the other authors are not available.

\listoffigures
\listoftables

\end{spacing}
\end{document}